\documentclass[useAMS,usenatbib]{mn2e}
\usepackage{graphicx}
\usepackage{textcomp}
\usepackage{txfonts}
\usepackage{comment}
\usepackage{capt-of}
\usepackage[hyperindex,breaklinks=true,colorlinks,citecolor=blue]{hyperref}
\usepackage{subfigure}
\usepackage{bm}
\usepackage[titletoc]{appendix}

\newcommand{\be}{\begin{equation}}
\newcommand{\ee}{\end{equation}}
\newcommand{\ba}{\begin{eqnarray}}
\newcommand{\ea}{\end{eqnarray}}

\newcommand{\nn}{\nonumber \\}

\newcommand{\C}{\mbox{\boldmath $C$}}

\newcommand{\rgl}{\rangle} 
\newcommand{\lgl}{\langle}
 
\newcommand{\ld}{{\ell '}}
 
\newcommand{\lm}{{\ell m}}
\newcommand{\lmd}{{\ell' m'}}

\newcommand{\lld}{{\ell \ell'}}

\newcommand{\sYlm}{\,_sY_\lm}
\newcommand{\sYlmd}{\,_sY_\lmd} 

\newcommand{\tYlm}{\,_2\!Y_\lm} 
\newcommand{\mtYlm}{\,_{-2}\!Y_\lm}

\newcommand{\lsim}{\mathrel{\rlap{\lower4pt\hbox{\hskip1pt$\sim$}}
    \raise1pt\hbox{$<$}}}                
\newcommand{\gsim}{\mathrel{\rlap{\lower4pt\hbox{\hskip1pt$\sim$}}
    \raise1pt\hbox{$>$}}}
\newcommand{\vgamma}{\bmath{\gamma}}

\newcommand{\vOmega}{\bmath{\Omega}}

\def\healpix{{\sevensize HEALPIX\,\,}}

\title[Cross-correlation cosmic shear with the SDSS and VLA FIRST
  surveys]{Cross-correlation cosmic shear with the SDSS and VLA FIRST surveys}
\author[Demetroullas \& Brown]{C.~Demetroullas$^{1}$\thanks{\href{mailto:constantinos.demetroullas@postgrad.manchester.ac.uk}{\nolinkurl{costantinos.demetroullas@postgrad.manchester.ac.uk}}} \& M.~L.~Brown$^{1}$\thanks{\href{mailto:m.l.brown@manchester.ac.uk}{\nolinkurl{m.l.brown@manchester.ac.uk}}}\\
$^{1}$Jodrell Bank Centre for Astrophysics, School of Physics and
Astronomy, The University of Manchester, Manchester, M13 9PL, UK. \\
}

\begin{document}
\pdfpageheight 11.692in 
\date{Accepted 2015 XXXXX XX. Received 2015 XXXXX XX; in original form 2015 XXXXX XX}

\pagerange{\pageref{firstpage}--\pageref{lastpage}} \pubyear{2015}

\maketitle

\label{firstpage}

\begin{abstract}
We measure the cosmic shear power spectrum on large angular scales by
cross-correlating the shapes of $\sim\!\!9$ million galaxies measured
in the optical SDSS survey with the shapes of $\sim\!\!2.7\times10^5$
radio galaxies measured by the overlapping VLA-FIRST survey.  Our
measurements span the multipole range $10 < \ell < 130$, corresponding
to angular scales $2^\circ < \theta < 20^\circ$. On these scales, the
shear maps from both surveys suffer from significant systematic
effects that prohibit a measurement of the shear power spectrum from
either survey alone.  Conversely, we demonstrate that a power spectrum
measured by cross-correlating the two surveys is unbiased, reducing
the impact of the systematics by at least an order of magnitude.

We measure an $E$-mode power spectrum from the data that is
inconsistent with zero signal at the 99\% confidence ($\sim2.7\sigma$)
level. The odd-parity $B$-mode signal and the $EB$ cross-correlation
are both found to be consistent with zero (within 1$\sigma$). These
constraints are obtained after a careful error analysis that accounts
for uncertainties due to cosmic variance, random galaxy shape noise
and shape measurement errors, as well as additional errors associated
with the observed large-scale systematic effects in the two surveys.
This latter source of uncertainty is particularly important for our
analysis as it amplifies the errors by a factor $\sim\!\!2.5$ compared
to the errors due to cosmic variance, galaxy shape noise and
measurement errors alone.  Our constraints, which probe the power
spectrum in the linear r\'egime, are consistent with the
expected signal in the concordance cosmological model assuming recent
estimates of the cosmological parameters from the \emph{Planck}
satellite, and literature values for the median redshifts of the SDSS
and FIRST galaxy populations.

The cross-power spectrum approach described in this paper represents a
powerful technique for mitigating shear systematics and will be ideal
for extracting robust results, with the exquisite control of
systematics required, from future cosmic shear surveys with the 
SKA, LSST, \emph{Euclid} and \emph{WFIRST-AFTA}.
\end{abstract}


\section{Introduction}\label{sec:intro}
Gravitational lensing is the apparent distortion in the shapes of
distant background galaxies due to the path of their light rays being
altered by the gravitational field of a foreground object. In the weak
lensing (WL) regime these distortions are so slight that they cannot
be measured using only one background source; hence WL is an
intrinsically statistical measurement. WL is considered today to be a
uniquely powerful tool in astronomy. This is because it can be used to
estimate the total mass of a foreground lens regardless of the
constituent matter's fundamental nature.

The tiny coherent WL distortions of background galaxy shapes caused by
the large-scale structure in the Universe is termed cosmic shear. The
theoretical basis for cosmic shear was pioneered by \citet{gunn1967},
but due to its challenging observational nature it was only detected
around three decades later \citep{Villumsen1995,Schneider1998,
  bacon2000,kaiser2000,waerbeke2000,wittman2000}. The field has since
flourished with numerous experiments delivering results with
increasing accuracy
\citep{vanwearbeke2001,brown2003,bacon2003,fu2008,heymans2012,jee2013}.
Measuring these gravitationally induced distortions at different
scales provides a wealth of cosmological information. In particular
cosmic shear studies can constrain the clustering amplitude $\sigma_8$ and the
matter density $\Omega_m$
\citep{hoekstra2006,schrabback2010,kilbinger2013}.  Furthermore with
forthcoming and future large surveys containing redshift information
(e.g. DES\footnote{Dark Energy Survey,
  \url{http://www.darkenergysurvey.org}}, KIDS\footnote{KIlo Degree
  Survey, \url{http://kids.strw.leidenuniv.nl}}, HSC\footnote{Hyper
  Suprime-Cam, \url{http://www.naoj.org/Projects/HSC}}
LSST\footnote{Large Synoptic Survey Telescope,
  \url{http://www.lsst.org}}, \emph{WFIRST-AFTA}\footnote{Wide-Field
  IR Survey Telescope, \url{http://wfirst.gsfc.nasa.gov}},
\emph{Euclid}\footnote{\emph{Euclid} satellite,
  \url{http://sci.esa.int/euclid}} and SKA\footnote{Square Kilometre
  Array, \url{http://www.skatelescope.org}}), it will be possible to
measure the evolution of the large-scale matter distribution and hence
place constraints on the dark energy equation of state $w$.

Nearly all cosmic shear studies to date have been performed in the optical and
near infrared (NIR). The only comparable study in the radio band is
the analysis of the VLA FIRST\footnote{Very Large Array; Faint Images
  of the Radio Sky at Twenty centimetres} survey described in
\cite{chang2004}. Weak lensing in the radio is still at an early stage
compared to lensing at optical/NIR wavelengths. The primary reason for
this is the relatively small number of sources in the radio sky at
current telescope sensitivities. Current radio surveys can deliver
source counts of $\sim$100\,deg$^{-2}$, of which $\sim$$25\%$ can be
used in weak lensing studies \citep{becker1995,chang2004}. This can be
compared to optical surveys which can reach source number density of
$\sim$10\,arcmin$^{-2}$ over similar area of sky
(e.g.~\citealt{ahn2014}). This, together with the inability of radio surveys to
deliver redshifts for the detected galaxies, means that WL studies in
the radio cannot currently compete with state-of-the-art optical/NIR
WL experiments.

This situation is expected to change in the foreseeable future. The
number counts of radio sources will increase dramatically when radio
telescopes breach the 1--10$\mu$Jy sensitivity level, at which point radio
and optical galaxy counts will be comparable. Ongoing and future
surveys with the JVLA\footnote{Jansky VLA,
  \url{https://science.nrao.edu/facilities/vla}} and
e-MERLIN\footnote{\url{http://www.e-merlin.ac.uk}} facilities, the SKA pathfinders
MeerKAT\footnote{\url{http://www.ska.ac.za/meerkat/}} and
ASKAP\footnote{\url{http://www.atnf.csiro.au/projects/askap/index.html}}, 
and finally the SKA itself, will achieve and
ultimately surpass this sensitivity level. Additionally with
the advent of the SKA, the detection of radio galaxies will also be
accompanied by estimates of their redshifts from measuring their HI
emission.

Where galaxy numbers are comparable, radio WL studies provide
potential advantages over their optical/NIR counterparts. Radio
interferometers have well-known and deterministic beam
patterns. Therefore the instrumental point spread function (PSF) can
be estimated, at any position on the sky, to very high accuracy. This
is in contrast to optical studies where the PSF is typically mapped
across the sky using point sources (stars) and is then extrapolated to
the positions of the galaxies. This PSF measurement and interpolation
scheme will inevitably introduce errors to the subsequent galaxy shape
measurements at some level.

Future radio surveys are also expected to be sensitive to the
star-forming galaxy population to higher redshifts than is typically
probed with optical surveys \citep{brown2013, brown2015}. Radio WL
surveys are therefore expected to probe the Universe at earlier
times. Additionally, \citet{brown2011} demonstrated the potential use
of polarisation information (which comes relatively easy with radio
observations) to estimate the orientations of galaxies prior to
lensing. This technique can be used to suppress the impact of
intrinsic galaxy alignments, a key astrophysical systematic effect
that contaminates weak lensing studies at all wavelenghts (see
\citealt{joachimi2015}, \citealt{kiessling2015} and \citealt{kirk2015}
for a recent overview). Radio polarization measurements can
potentially also be used to significantly reduce the number of
galaxies required for a weak lensing study. Looking to even higher
redshifts, \citet{pourtsidou2014} have shown that one can exploit 21cm
SKA surveys to perform weak lensing studies without having to resolve
or even identify individual galaxies. For more details on the SKA WL
prospects in general see \citet{brown2015}.

Another key advantage, which we aim to exploit in this paper, arises
when radio and optical/NIR data are combined. WL analyses rely on
extremely accurate (and, to a lesser extent, precise) measurements of
galaxy shapes. However instrumental systematic effects can introduce
correlated errors in the measurements. Since it is primarily
correlations in the galaxy shapes that one wishes to measure for
cosmology, such instrumental systematics represent a significant
problem for WL surveys. The source of systematic effects in
optical/NIR telescopes and radio interferometers are of a very
different origin and nature, and are therefore not expected to
correlate. By cross-correlating the shape information from two such
experiments, one therefore expects to cancel out any instrumental
systematics and obtain an unbiased cosmic shear measurement \citep{jarvis2008,
  patel2010}. In this paper, we aim to demonstrate the potential
benefits of such cross-correlaton techniques by performing an
exploratory optical-radio cross-correlation weak lensing analysis of
the VLA FIRST survey \citep{becker1995} and the Sloan Digital Sky
Survey (SDSS, \citealt{york2000}). These two surveys are well suited
for this investigation as they provide approximately 10,000 deg$^2$ of
overlapping sky coverage. In contrast to most other cosmic shear analyses, our
measurements probe the shear power spectrum on large scales, at
multipoles $\ell < 100$, corresponding to angular scales above a few
degrees. We perform our analysis using a cross-power spectrum
approach, which provides for a natural decomposition of the signal
into its constituent $E-$ and $B-$modes, and also fully accounts for
the surveys' masks and geometries on the curved sky.

The paper is organised as follows. In Section~\ref{sec:science} we
present the relevant weak lensing formalism, and the details of our
algorithm for estimating the cosmic shear cross-power spectrum from
the data. Section~\ref{sec:data} describes the two surveys that we use
in the analysis. In Section~\ref{sec:syst_and_maps} we describe how we
construct shear maps from the two datasets. We identify large-scale
systematic effects in these maps which dominate the measured shear
auto-correlations in either dataset alone. We demonstrate using
simulations in Section~\ref{sec:sims} that the systematics can be
removed by cross-correlating the shape information from the two
datasets. In Section~\ref{sec:rr} we present the results from the
analysis of the real data where we measure the shear cross-power
spectrum from the FIRST and SDSS datasets. We discuss our results and
conclude in Section~\ref{sec:concl}.

\section{Weak lensing theory}\label{sec:science}
Distortions in the shapes of distant galaxies can be quantified via
measurements of their ellipticity, which is a spin-2 quantity:
\begin{equation}
 \epsilon_{ij}=  \left ( \begin{tabular}{cc} $ \epsilon_1 $ &  $
   \epsilon_2 $  \\ $ \epsilon_2 $ & $ -\epsilon_1 $  \end{tabular}
 \right ).
 \label{eq:ellip_def}
\end{equation}
The two components of this matrix represent the two orthogonal modes
of an ellipse: $\epsilon_1$ describes elongation or compression along
or perpendicular to a user-defined reference axis while $\epsilon_2$
describes elongation or compression at an angle $\pi/4$ to the
reference axis. Galaxy ellipticity can also be described in terms of a
semi-major axis ($a$), a semi-minor axis ($b$), and a position angle
($\phi$) as \ba \epsilon_1&=&\epsilon \times
\cos(2\phi),\label{eq:majmintoe1} \\ \epsilon_2&=&\epsilon \times
\sin(2\phi).\label{eq:majmintoe2} \ea Here, $\epsilon = (a-b) / (a+b)$
is the modulus of the galaxy's ellipticity and $\phi$ is the position
angle of the galaxy with respect to the reference axis of the chosen
coordinate system.

In the absence of intrinsic galaxy alignments, galaxy ellipticity is
an unbiased estimator for the lensing shear field, which is also a
spin-2 quantity,
\begin{equation}
\lgl \epsilon_{ij} \rgl = \gamma_{ij} \equiv  \left( 
\begin{tabular}{cc} $ \gamma_1 $ &  $ \gamma_2 $  \\ 
$ \gamma_2 $ & $ -\gamma_1 $  \end{tabular} \right),
 \label{eq:ellip_expval}
 \end{equation}
where the angled brackets denote an ensemble average. Since gravity is
a potential theory, the shear at angular position $\Omega$ can be
related to a lensing potential ($\psi$) as
\begin{equation}
\gamma_{ij}(\Omega) = \left( \delta_i \delta_j -\frac{1}{2} \delta^K_{ij}
\delta^2 \right) \psi(\Omega),
\label{eq:shear_def}
\end{equation}
where $\delta_i \equiv r (\delta_{ij} - \hat{r}_i \hat{r}_j \nabla_i)$
is a dimensionless, transverse differential operator, and $\delta^2 =
\delta_i \delta^j$ is the transverse Laplacian. The lensing potential
can in turn be related to the 3-D gravitational potential,
$\Phi(\mathbf{r})$ by (e.g.~\citealt{kaiser1998}) 
\begin{equation}
\psi(\Omega) = \frac{2}{c^2} \int^r_0 dr'
\left(\frac{r-r'}{rr'}\right)\Phi(\mathbf{r'}),
\end{equation}
where $r$ is the comoving distance to the sources. 

Note that the ellipticity and shear fields of
equations~(\ref{eq:ellip_def})--(\ref{eq:ellip_expval}) are defined with
respect to an arbitrarily defined reference axis. In the case where
one is interested in the weak lensing distortion in a population of
background sources due to the presence of a known foreground object
(or ``lens''), it is more natural to consider the tangential and
rotated shear (or ellipticity), defined by
\ba
\gamma_t&=&\gamma_1 \cos(2\theta) + \gamma_2
\sin(2\theta), \label{eq:gammat} \\
\gamma_r&=&-\gamma_1 \sin(2\theta) + \gamma_2 \cos(2\theta),
\label{eq:gammar}
\ea
where $\theta$ is the position angle formed by moving counter
clockwise from the reference axis to the great circle connecting each
source-lens pair. The tangential shear, $\gamma_t$, describes
distortions in a tangential and/or radial direction with respect to
the lens position. The rotated shear, $\gamma_r$, describes
distortions in the orthogonal direction, at an angle $\pm \pi/4$ to
the vector pointing to the lens position.

As indicated by equation~(\ref{eq:ellip_expval}), the shear components can
be estimated by averaging over a set of observed background galaxy
ellipticities. The shear field can itself be related to the lensing
convergence field (which measures the projected surface mass density
contrast) by \citep{kaiser1993}
\begin{equation}
\kappa=\partial ^{-2} \partial_i \partial_j \gamma_{ij},
\label{eq:kappa}
\end{equation}
where $\partial ^{-2} $ is the inverse 2-D Laplacian operator defined by 
\begin{equation}
\partial ^{-2} \equiv \frac{1}{2\pi} \int d^2\hat{r}' \ln \left|\hat{r}-\hat{r}'\right|.
\end{equation}
A second decomposition of the shear yields the odd-parity divergence
field,
\begin{equation}
\beta = \partial^{-2} \varepsilon_i^{n} \partial_j \partial_n \gamma_{ij},
\label{eq:beta}
\end{equation}
where $\varepsilon_i^n$ is the Levi-Civita symbol in two dimensions,
\be
\varepsilon_i^n = \left ( \begin{tabular}{cc} 0 &  -1  \\ 1 & 0  \end{tabular} \right ).
\ee
Since gravitational lensing produces no $\beta$-modes, at least to first
order, the divergence field is a useful quantity for testing
systematic effects in weak lensing studies.

\subsection{Weak lensing in spherical harmonic space}\label{sec:fourier_lensing}
As mentioned above, lensing shear is a spin-2 field, i.e. it
transforms as $\vgamma \rightarrow \vgamma e^{i2\phi}$ under rotation
by $\phi$ (where, for convenience we have defined the complex shear
$\vgamma = \gamma_1 + i \gamma_2$). One may thus expand $\vgamma$ and
its complex conjugate in terms of the spin-weighted spherical
harmonics, $\sYlm$ \citep{newman1966} as
\ba 
\vgamma (\Omega)  &=& \gamma_1 (\Omega) + i \gamma_2 (\Omega) \nn
	    &=& \sum_\lm (\kappa_\lm + i \beta_\lm)\tYlm(\Omega), \\
\vgamma^*(\Omega)  &=& \gamma_1 (\Omega) - i \gamma_2 (\Omega) \nn
	    &=& \sum_\lm (\kappa_\lm - i \beta_\lm)\mtYlm(\Omega), 
\ea 
where $s$ denotes the spin and the summation in $m$ is over $-\ell \le
m \le \ell$. $\kappa_\lm$ and $\beta_\lm$ are the spin-2 harmonic modes of the
so-called electric (i.e. ``gradient'' or ``$E$-mode'') and magnetic
(i.e. ``curl'' or ``$B$-mode'') components of the shear field
respectively, which we further identify as the harmonic space versions
of the lensing convergence and divergence fields mentioned in the
previous section. Using the orthogonality of the spin-s spherical
harmonics over the whole sphere,
\ba \int \! d \Omega\, \sYlm(\Omega) \sYlmd^{*}(\Omega) = \delta_{\ell
  \ld} \delta_{m m'}, 
\ea 
where the spin states must be equal, the
harmonic modes of the $\kappa$ and $\beta$ fields can be found
directly from the shear field, $\vgamma$; 
\ba \kappa_\lm &=&
\frac{1}{2} \int \! d \Omega\, [\vgamma (\Omega)\tYlm^*(\Omega) +
  \vgamma^* (\Omega)\mtYlm^*(\Omega)], \label{eq:kappa_lm}\\ \beta_\lm
&=& \! \frac{-i}{2} \! \int \! d \Omega\,[\vgamma(\Omega)
  \tYlm^*(\Omega) - \vgamma^*(\Omega)
  \mtYlm^*(\Omega)].
\label{eq:beta_lm} 
\ea
Taking the average values over the sphere of products of the
harmonic coefficients of the $\kappa$ and $\beta$ fields, one can
construct three possible power spectra:
\ba
C_\ell^{\kappa\kappa} = \frac{1}{2\ell+1} \sum_m  \kappa_\lm \, \kappa^*_\lm, \label{eq:clkk_est}\\
C_\ell^{\beta\beta} = \frac{1}{2\ell+1} \sum_m  \beta_\lm \, \beta^*_\lm, \label{eq:clbb_est}\\
C_\ell^{\kappa\beta} = \frac{1}{2\ell+1} \sum_m  \kappa_\lm \, \beta^*_\lm. \label{eq:cl_kb_est} 
\ea
The parity invariance of weak lensing suggests that $C_\ell^{\beta\beta}$
and $C_\ell^{\kappa\beta}$ should be equal to zero in the absence of
systematics. However, systematic effects (both instrumental and
astrophysical) can give rise to a non-zero $C_\ell^{\beta\beta}$.  Finite
fields and boundary effects can also lead to leakage of power
between the three spectra.

In the limit of weak lensing, the two-point statistical properties of
the shear and convergence fields are the same \citep{blandford1991} so
that $C_\ell^{\gamma \gamma}=C_\ell^{\kappa \kappa}$. Finally, we can relate
the convergence power spectrum to the 3-D matter power spectrum
$P_{\delta}(k,r)$ through (e.g.~\citealt{bartelmann2001b})
\begin{equation}
C_\ell^{\kappa(i)\kappa(j)}=\frac{9}{4} \left( \frac{H_0}{c} \right)^4
\Omega^2_m \int ^{r_H}_0 dr \, P_{\delta}\left(
\frac{l}{r},r\right)\left(\frac{\overline{W}_i(r)\overline{W}_j(r)}{a(r)^2} \right),
\label{eq:clkk_theory}
\end{equation}
where we have assumed a flat Universe, and we have now considered the
more general case where one measures the cross-correlation signal
between two different surveys, here denoted with indices $i$ and $j$. In
equation~(\ref{eq:clkk_theory}), $H_0$ is the Hubble constant, $a$ is
the scale factor of the Universe, $r$ is comoving distance, $r_H$ is
the comoving distance to the horizon and $\Omega_m$ is the matter
density. The weighting, $\overline{W}_i(r)$, is given in terms of the
normalised source distribution for each survey, $G_i(r)dr = p_i(z)dz$:
\begin{equation}
\overline{W}_i(r)\equiv \int ^{r_H}_r dr' \, G_i(r')\frac{r' - r}{r'}.
\end{equation}
 
\subsection{Power spectrum estimation on a cut sky}\label{sec:pseudo_cl_theory}
The power spectrum estimators of
equations~(\ref{eq:clkk_est})--(\ref{eq:cl_kb_est}) are only unbiased in the
case of full sky coverage and in the absence of noise. However, in the analysis
which follows we shall be estimating these power spectra from noisy
data covering only a fraction of the sky. In order to correct for the
effects of limited sky coverage and noise, we adopt a
``pseudo-$C_\ell$'' analysis, originally developed in the context of CMB
polarization observations. We provide a brief summary of the technique
here and refer the reader to \cite{brown2005} for further details. 

In the presence of finite sky coverage and/or a mask that excludes
certain regions of the sky, one can estimate the so-called
pseudo-$C_\ell$'s directly from a set of shear maps (using fast spherical
harmonic transforms from e.g.~the \healpix software\footnote{see
  \url{http://healpix.sourceforge.net}}, \citealt{gorski2005}) as:
\be
\widetilde{C}^{A(i)B(j)}_\ell = \frac{1}{2\ell+1} \sum_m \tilde{A}_\lm(i) \tilde{B}^*_\lm(j),
\label{eq:cross-power-est}
\ee
where $\{A,B\}$ are any of the three possible combinations of $\kappa$
and $\beta$ and $\{i,j\}$ denote the two different surveys as
before. In the following sections, we will mostly be interested in
taking the cross-power spectra of the FIRST and SDSS datasets,
i.e. where $i$ denotes the FIRST dataset and $j$ denotes SDSS. When
cross-correlating two datasets, there are then in principle two
possible $C_\ell^{\kappa\beta}$ power spectra. For simplicity, we will
take the average of these and present a single $C_\ell^{\kappa\beta}$
cross-power spectrum. In equation~(\ref{eq:cross-power-est}),
$\tilde{A}_\lm$ and $\tilde{B}_\lm$ denote the spin-2 spherical
harmonic modes (found using equations~\ref{eq:kappa_lm} and
\ref{eq:beta_lm}) of weighted versions of the lensing shear field, 
\be
\widetilde{\vgamma}(\Omega) = W(\Omega) \vgamma (\Omega),
\label{eq:weighted_shear_field}
\ee
where $W(\Omega)$ is an arbitrary weighting function which can be used to
exclude certain regions of the sky and/or downweight noisier sky
pixels.

Grouping all three power spectra into a single vector,
$\C_\ell = \{C_\ell^{\kappa\kappa}, C_\ell^{\beta\beta},
C_\ell^{\kappa\beta}\}$, one can show that the pseudo-$C_\ell$'s
measured from the weighted shear fields are related to the true power
spectra on the full sky via
\be
\widetilde{\C}_\ell = \sum_\ld M_\lld \C_\ld,
\label{eq:pseudo-cl-expval}
\ee
where we have introduced the coupling matrix, $M_\lld$. This coupling
matrix fully encodes how the survey geometries and masks (described
via the function, $W(\Omega)$) mix modes, both within a single
spectrum (e.g. $C_\ell^{\kappa\kappa} \rightarrow
C_\ld^{\kappa\kappa}$), and also between spectra
(e.g. $C_\ell^{\kappa\kappa} \rightarrow C_\ld^{\beta\beta}$), and it
can be calculated exactly for any given set of weighting functions,
$W(\Omega)$. Specifically, in our case where we are estimating the
cross-power spectra between two datasets, we can calculate the
$M_\lld$ matrix exactly from the pseudo-$C_\ell$ cross-power spectra
(estimated using equation~\ref{eq:cross-power-est}) of the
weighting functions used for the two different surveys, $W_i(\Omega)$
and $W_j(\Omega)$. Further details and explicit formulae for calculating
$M_\lld$ are provided in \cite{brown2005}.

Recovering an estimate of the power spectra on the full sky is then
only a matter of inverting equation~(\ref{eq:pseudo-cl-expval}). When
calculating auto-power spectra (from a single survey), one also needs
to include a correction for the noise bias arising from the intrinsic
dispersion in galaxy shapes and measurement errors. For practical
reasons, it is often also convenient to estimate the power spectrum in
terms of ``band powers'' which recover the average power across a range of
multipoles, $\Delta\ell$. Including these features, the power spectrum
estimator becomes 
\begin{equation}
\hat{P}_b=\sum_{b'} K^{-1}_{bb'}\sum_{\ell}O_{b'\ell} (\widetilde {C_\ell} -
\lgl \widetilde {N_\ell} \rgl_{\rm MC}),
\label{eq:bandpower_est}
\end{equation}
where $\hat{P}_b$ denotes the debiased combined estimator for
band powers of the $C_\ell^{\kappa\kappa}$, $C_\ell^{\kappa\beta}$ and
$C_\ell^{\beta\beta}$ power spectra. In
equation~(\ref{eq:bandpower_est}), $\widetilde{\C_\ell}$ are the
pseudo-$C_\ell$'s measured from the data and
$\lgl\widetilde{N_\ell}\rgl$ are the noise-only pseudo-$C_l$'s
estimated from a suite of Monte Carlo (MC) simulations which contain
realisations of the noise, and also incorporate the effects of the survey
masks. Note that, for our cross-power spectrum analysis, the noise is
uncorrelated and so the noise subtraction step (subtraction of the $\lgl
\widetilde{N_\ell} \rgl_{\rm MC}$ term) is not required, although we
have nevertheless implemented it in our analysis pipeline for
completeness.  

$K_{bb'}$ is the binned coupling matrix, constructed from $M_\lld$ as
\begin{equation}
K_{bb'}=\sum_\ell O_{b\ell} \sum _{\ld} M_{\lld} F_\ld Q_{\ld b'},
\end{equation}
where $O_{b\ell}$ is a binning operator that bins the $C_\ell$'s into
band powers and $Q_{\ell b}$ is the inverse operator which ``unfolds''
the band powers into individual $C_\ell$'s. The function, $F_\ell$
describes the smoothing effect on the underlying shear fields due to
the fact that the sky maps are pixelized. This function is provided by
the {\sevensize HEALPIX} software that we use to pixelize the sky and
perform the spin spherical harmonic transforms.  

The binning operator we choose to use is  
\begin{equation}
O_{b\ell} = \left\{ \begin{array}{l l}
	\frac{1}{2\pi}\frac{\ell(\ell+1)}{\ell_{low}^{(b+1)}-\ell_{low}^{(b)}}~,
        & \quad \mathrm{ if}~ 2 \leq \ell_{low}^{(b)} \leq \ell <
        \ell_{low}^{(b+1)}\\
	0~, & \quad \mathrm{otherwise}~, \end{array} \right.
\end{equation}
where $\ell_{low}^{(b)}$ denotes the lower edge of band $b$. With this
choice, the quantity $\ell(\ell+1)C_\ell/2\pi$ is approximated as
constant within each band power. 

\section{The surveys}\label{sec:data}

\subsection{FIRST data}\label{sec:first_data}
The VLA FIRST survey \citep{becker1995} is a 10,575 deg$^2$ survey of
the sky in the declination range $-10^{\circ} \leq \delta \leq
70^{\circ}$, and right ascension range $8 \, {\rm hrs} \leq RA \leq 17
\, {\rm hrs}$. Of the survey's total sky coverage, 8,444 deg$^2$ are
in the north galactic cap and 2,131 deg$^2$ are in the south. The data
were gathered using the VLA at L-band (1.4~GHz), in B-configuration
(maximum baseline length of 11.1 km) and in snapshot mode ($<$5
minutes per observation). The observations were conducted between 1993
and 2011. The coverage for the southern cap is discontinuous due to
poor weather and system failures during the 2011 observations.

The beam of the telescope at $\delta > +4^{\circ}33'$ can be
approximated by a circular Gaussian with a full width at half maximum
(FWHM) of 5.4''. Below $\delta = +4^{\circ}33'$ the beam is elliptical
with a FWHM of 6.4''$\times$5.4'', with the major axis running
north-south. For $21 \, {\rm hrs} < RA < 3 \, {\rm hrs}$ and $\delta <
-2^{\circ}30'$ the ellipticity of the beam increases further to
6.8''$\times$5.4''. The survey achieved an RMS noise level of
$\sim$0.15 mJy, and delivered positions, flux densities and shape
information for $\sim$1 million radio sources, of which $\sim$40\% are
resolved. Shape information was extracted by fitting an elliptical
Gaussian model to each source. The FIRST catalogue provides this shape
information via a major and minor axis and a position angle (PA) for
each object. Note that the FIRST analysis pipeline includes a
deconvolution step to remove blurring due to the telescope's PSF. More
information regarding the contents of the FIRST catalogue and how it
was generated is available from the FIRST 
website\footnote{\label{fw}FIRST website: \url{http://sundog.stsci.edu/index.html}}.

Mainly due to the relatively sparse coverage in visibility (or
``$uv$'') space -- an unavoidable result of the observation strategy
-- the deconvolution/{\sevensize CLEAN}ing\footnote{%
The {\sevensize CLEAN} algorithm \citep{hogbom1974} is a standard
technique commonly used in radio astronomy to deconvolve images for
the effects of a finite PSF. The algorithm models the data as a
collection of point sources and, starting with the brightest source,
iteratively subtracts (from the $uv$-data) the flux associated with
each source in order to detect and characterise fainter objects.}
of the FIRST sources was imperfect. The catalogue therefore contains
information about the possibility of a detected source being a
residual sidelobe of a nearby brighter source.  Of the total number of
FIRST sources, 76.4\% had a probability of being spurious, P(S), less
than 5\%. For 17\% of the sources the probability was between 5 and
50\%. The remaining 6.5\% had more than a 50\% chance of being a
sidelobe of a nearby bright galaxy. Additional information about the
sources was generated by cross-correlating their positions with
positions of optical sources from the SDSS and 2-micron All-Sky Survey
(2MASS, \citealt{skrutskie2006}) catalogues.

The FIRST data have previously been used to make a detection of cosmic
shear in the radio band by extracting source shape information
directly from the $uv$ plane \citep{chang2004}. Those authors measured
the 2-point correlation function (2PCF) on angular scales, $1^{\circ}
< \theta < 40^{\circ}$. On scales $1^{\circ} < \theta < 4^{\circ}$ the
measured $\beta$-modes where consistent with zero while a significant
signal was detected in the $\kappa$-modes at the $\sim$3$\sigma$
level. From their results, \cite{chang2004} placed a joint constraint
on the clustering amplitude, $\sigma_8$ and the median redshift of the
FIRST sources without an optical counterpart, $z_m$ of
$\sigma_8(z_m/2)^{0.6}\simeq1.0\pm0.2$. Adopting a prior of
$\sigma_8=0.9\pm0.1$ they obtained the constraint $z_m = 2.2\pm0.9$
(68\% CL). This value is consistent with existing models of the radio
source luminosity function. The results are encouraging as they show
that although radio surveys lack the sensitivity at the moment to
compete with optical lensing surveys, they can deliver shape
information which, with the proper processing, can be used for weak
lensing studies.

For the purposes of this study we acquired the {\sevensize 13JUN05}
version of the FIRST catalogue. The \citet{chang2004} study estimated
the median redshift of the entire FIRST sample to be between 0.9 and
1.4 (depending on the model of the radio source luminosity function
used). Using the SKA simulated skies (S$^3$,\citealt{wilman2008}) -- a
computer simulation based on more recent data, and designed to model
the radio and submillimetre Universe -- we estimate the median
redshift of the entire FIRST sample to be $z_m\sim\!\!1.2$.

\subsection{SDSS data}\label{sdss_data}
The SDSS is an ongoing (since 2000) optical survey of the north and
south galactic caps north of declination -15$^{\circ}$, covering
$\sim$14,500 deg$^2$ of the sky. The survey uses a 2.5 metre telescope
located at Apache Point Observatory (APO) in the Sacramento mountains
in south New Mexico \citep{abazajian2003,ahn2014}. In 2008 the
experiment entered a new phase called SDSS-III in which new
instruments came into operation \citep{eisenstein2011}. One of the
SDSS-III surveys, the Baryon Oscillation Spectroscopic Survey (BOSS;
\citealt{dawson2013}), obtained spectra of $\sim$1.4 million luminous
red galaxies (LRGs) and $\sim$0.3 million quasars which have
subsequently been used to place stringent constraints on cosmological
models \citep{aubourg2014}.
 
The SDSS data have been used in numerous weak lensing experiments
probing different angular scales on the sky. The data have proved
particularly useful for galaxy-galaxy lensing
(e.g.~\citealt{fischer2000, mandelbaum2006}) and cluster lensing
(e.g.~\citealt{mandelbaum2008, sheldon2009, rozo2010}). In terms of
cosmic shear measurements, \cite{lin2012} and \cite{huff2014} have
both exploited the deep observations of SDSS ``Stripe 82'' to perform
cosmic shear analyses. Both of these latter studies detected a
significant $\kappa$-mode signal on degree scales with $\beta$-modes
found to be consistent with zero. Derived constraints on $\Omega_m$
and $\sigma_8$ were also found to be consistent between these two
independent studies.

In this study, we have used the tenth data release (DR10) of SDSS
\citep{ahn2014}. DR10 delivers photometric information across 5 bands
(u, g, r, i, z) for $\sim$500 million galaxies and stars (the PHOTO
sample), and spectra for $\sim$2.5 million of them (hereafter, the
SPECTRO sample). The mean magnitude across all PHOTO galaxies and
bands is $\sim$22.5. The median magnitude and redshift for the SPECTRO
galaxy population is $\sim$21 and $\sim$0.3 respectively. The SPECTRO
sample includes two distinct populations one of which peaks at $z
\approx 0.15$ and the second at $z \approx 0.5$. Only a few galaxies
in the sample are at redshifts greater than $\sim\!0.7$. We chose to
download all sources from the SDSS DR10 PHOTO sample that were
identified in i-band as galaxies with magnitudes i$_{\rm
  mag}<$\,26. For the $\sim$38.5 million sources that met the criteria
we acquired information about their positions ($RA$ and $\delta$) and
multi-band information describing their flux densities, galaxy types,
second and fourth moments, the reconstructed PSF at each galaxy
position, the galaxy magnitudes and quoted errors, and finally their
extinction.

\section{Shear maps and tests for systematics}\label{sec:syst_and_maps}
Telescope induced effects can severely bias a WL study. Therefore
systematics must be accounted for and if possible corrected. In this
section, we describe how we construct pixelized maps of the FIRST and
SDSS shear fields from the two galaxy catalogues, and we assess the
resulting maps for systematic effects. 

\subsection{FIRST analysis}\label{sec:first_syst_and_maps}
\subsubsection{Source selection}\label{sec:first_src_select}
A detailed study of the systematic effects in the FIRST survey was
performed in \citet{chang2004}. To control systematics,
\citet{chang2004} first performed a series of cuts on the source
catalogue in order to remove from the sample those sources whose
shapes were most likely to be corrupted due to residual systematic
effects. In this analysis, we have applied as many of their source
selection criteria as possible given the information that was
available to us in the FIRST catalogue. We therefore discarded sources
that had an unresolved deconvolved minor axis, or a deconvolved major
axis that was greater than 7 arcsec or smaller than 2 arcsec. We also
removed sources that had an integrated flux that was smaller than 1
mJy. Finally we have included only sources that had a possibility of
being a sidelobe of a nearby brighter source P(S) $< 5\%$. After
applying these cuts we were left with $\sim$2.7$\times10^5$
radio sources (from an initial sample of $\sim$1 million catalogue
entries). For all remaining sources, we formed the $\epsilon_1$ and
$\epsilon_2$ ellipticity estimates from the catalogued major and minor
axes, and position angle, according to equations~(\ref{eq:majmintoe1})
and (\ref{eq:majmintoe2}).

\subsubsection{Residual systematics in FIRST ellipticities}\label{sec:first_beam_sys}
\begin{figure} 
\centering
\includegraphics[width=8.5cm]{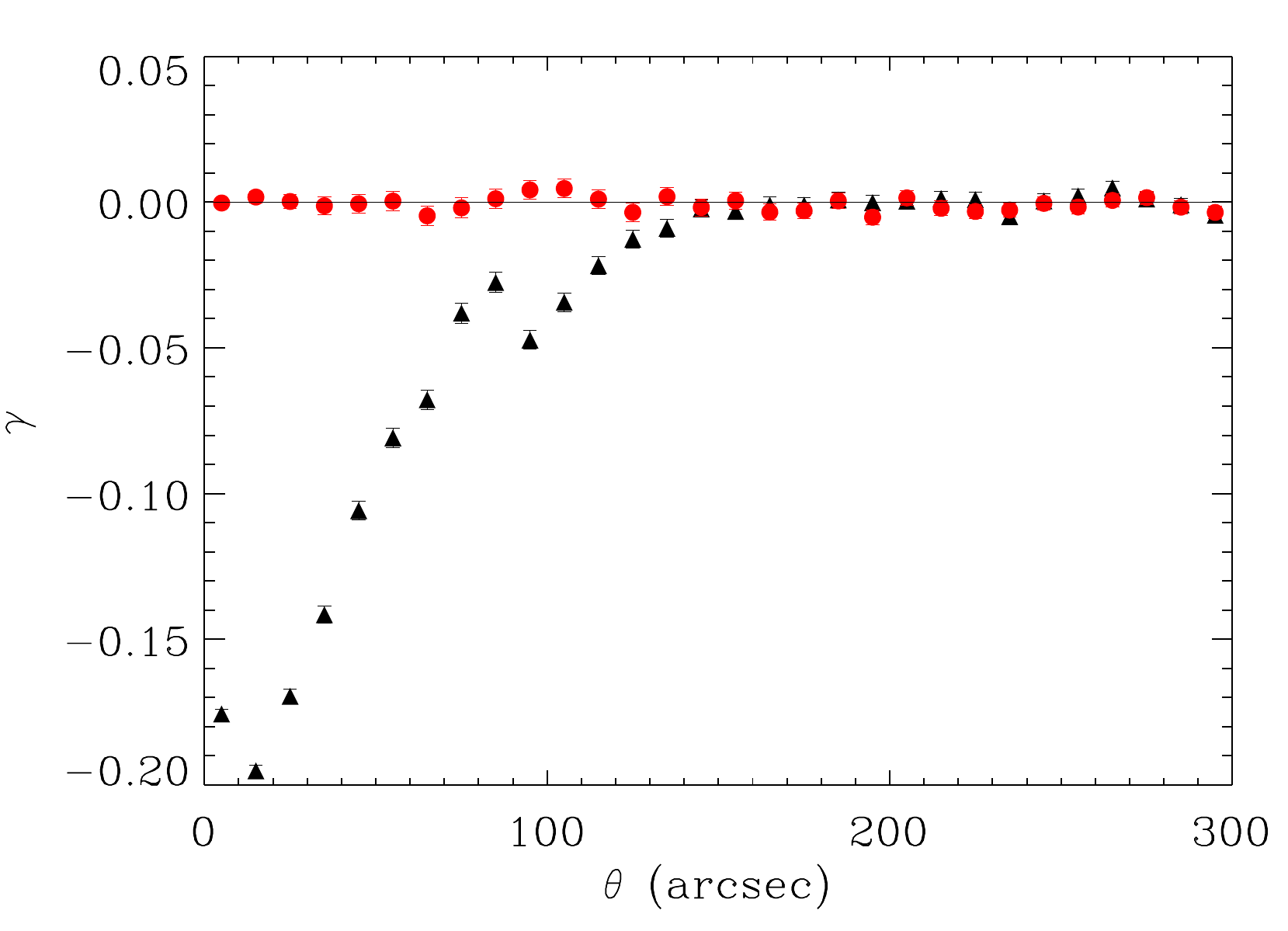} 
\caption{Results of the test for residual beam systematics in the
  FIRST galaxy shapes, as described in Section
  \ref{sec:first_beam_sys}. A large radial (negative $\gamma_t$,
  black triangles) distortion, decreasing as a function of angular
  scale, $\theta$, is found in the FIRST galaxy ellipticities when
  their shapes are stacked around the FIRST galaxy positions. The
  rotated shear ($\gamma_r$, red circles) is consistent with zero.}
\label{fig:contazim}
\end{figure}

We have performed a number of quality checks on the data immediately
after applying the cuts described above. By far the most informative
of these was a test for corruption of the galaxy ellipticities due to
residual sidelobe contamination resulting from an imperfect
deconvolution and/or {\sevensize CLEANING} procedure. To perform this
test, we have made use of the g-g lensing tangential and rotated shear
constructions ($\gamma_t$ and $\gamma_r$; equations~\ref{eq:gammat} \&
\ref{eq:gammar}) and have stacked the shapes of the selected FIRST
galaxies around the positions of all of the sources in the original
FIRST catalogue (including both resolved and unresolved objects).
The results of this test are shown in Fig.~\ref{fig:contazim} where we see
a strong distortion in the FIRST galaxy shapes oriented radially from
the central stacking position. This signal persists when we randomly
choose a sub-set of the FIRST sources' shapes and/or
positions. Moreover, the signal shows no obvious dependency on the
flux of the stacked and/or central sources.

\begin{figure*}
\subfigure{\includegraphics[width=7.5cm]{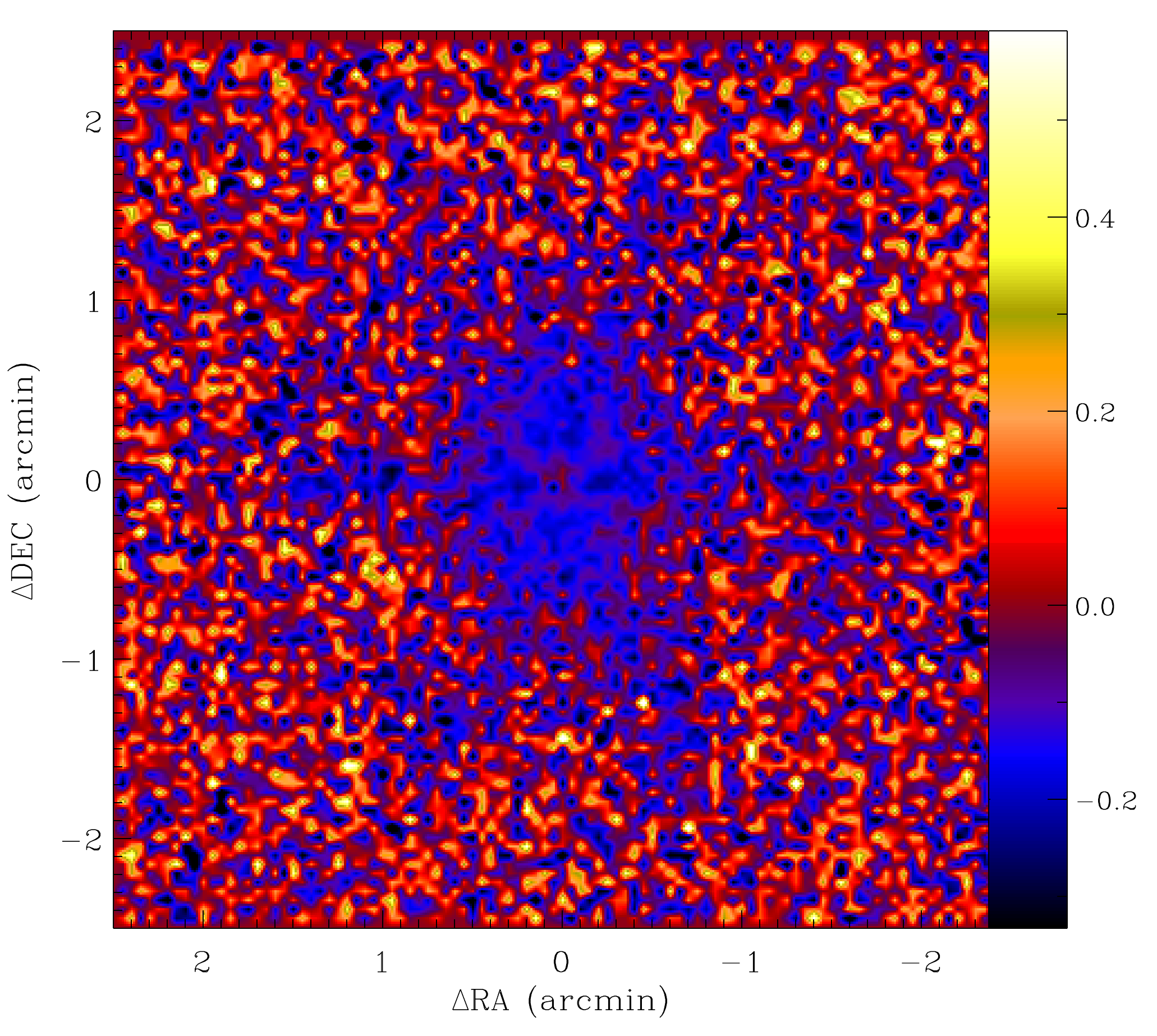}}\hspace{0.5em}
\subfigure{\includegraphics[width=7.5cm]{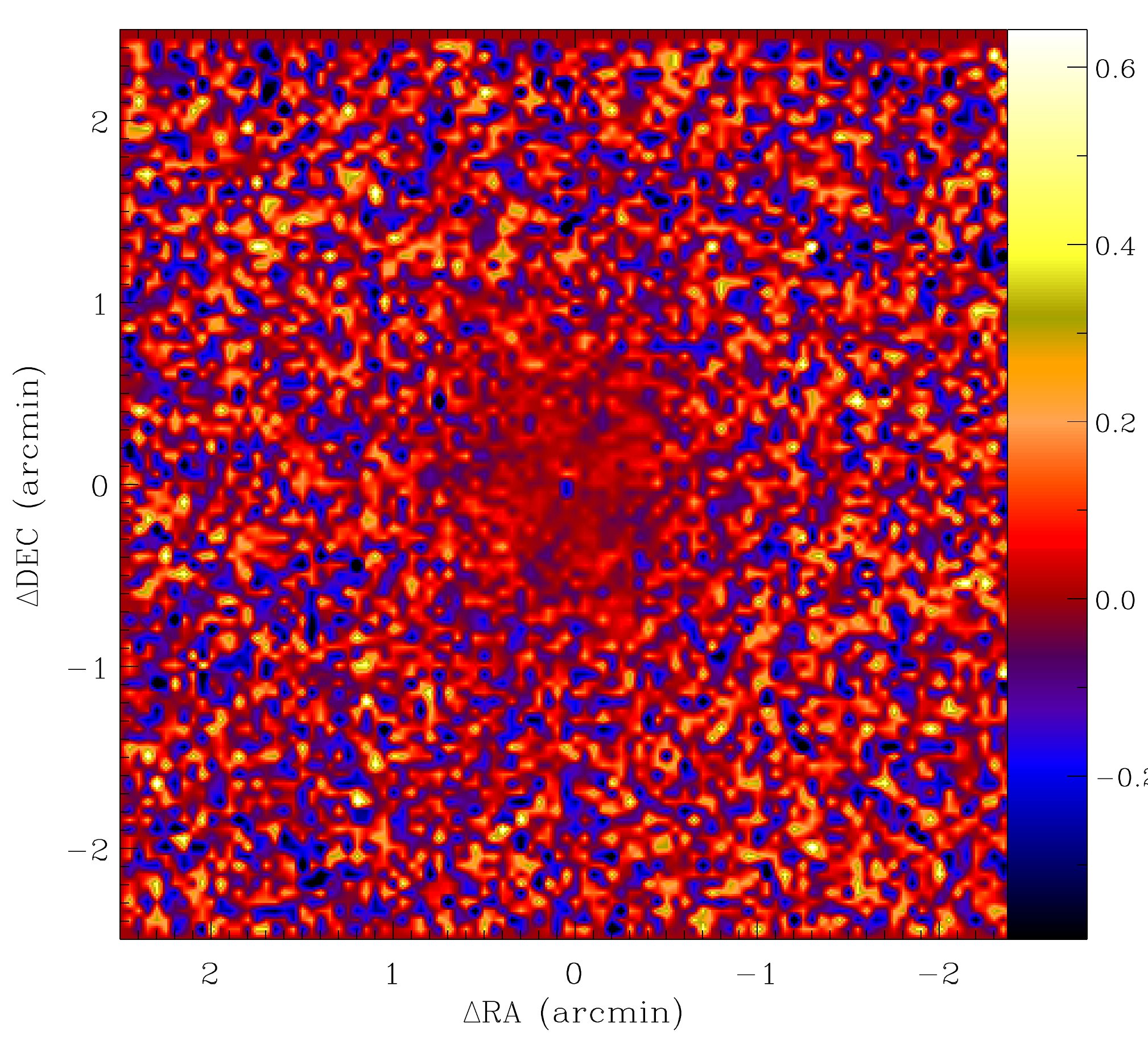}}
\caption{Maps of the tangential shear ($\gamma_t$, \emph{left panel}) and
  rotated shear ($\gamma_r$, \emph{right panel)} as a
  function of the separation in $RA$ and $\delta$ from the central
  stacking positions. The tangential shear map reveals a negative
  amplitude 6-arm star pattern which closely resembles the structure
  in the synthesised beam (PSF) of the VLA snapshot observations.}
\label{fig:contdraddec}
\end{figure*}

To further investigate the origin of this signal we repeated the
stacking analysis in ($\Delta RA, \Delta \delta$) space (i.e. where the
positions of all of the central sources are re-mapped to $\Delta RA = 0;
\Delta \delta = 0$). The resulting maps of tangential and
rotated shear are shown in Fig.~\ref{fig:contdraddec}. 
The 6-arm star pattern apparent in the $\gamma_t$ map, displayed in
the left panel of Fig.~\ref{fig:contdraddec}, closely matches the
synthesised beam (or PSF) of the VLA in ``snapshot mode'' which was the
observation mode employed during collection of the FIRST survey
data. The results of this test clearly indicate the presence of
residual systematics in the FIRST galaxy ellipticities, which are
strongly correlated with the known PSF of the VLA-FIRST
observations. These residuals are clearly spurious, and are almost
certainly the result of an imperfect deconvolution and/or {\sevensize
  CLEANING} of the FIRST data during the imaging step in the FIRST
data reduction. In the analysis which follows, we will attempt to
include the effects of these residual ellipticity correlations by
incorporating a model of the systematic in our MC
simulations. However, we also expect our cross-power spectrum approach
-- whereby we cross-correlate the FIRST and SDSS galaxy shapes -- to
be robust to the presence of this FIRST-specific systematic effect.

\subsubsection{FIRST shear maps}\label{sec:first_shear_maps}
The pseudo-$C_\ell$ power spectrum analysis outlined in
Section~\ref{sec:pseudo_cl_theory} requires as input pixelized maps of
the shear components. We create these maps by simple averaging of the
ellipticities of the FIRST sources which fall within each sky
pixel and we use the {\sevensize HEALPIX} scheme \citep{gorski2005} to
define the sky pixelisation. We choose the pixel size according to the
density of sources in the FIRST survey: with  $\sim$\,2.7$\times10^5$
sources surviving the cuts described in
Section~\ref{sec:first_src_select}, a pixel size of $\sim$1 deg$^2$
results in a mean pixel occupation number of $\sim$20 galaxies. This
choice ensures that our maps are contiguous over the survey area and
are mostly free from unoccupied pixels within the survey
boundaries. We therefore set the {\sevensize HEALPIX} resolution
parameter, $N_{\rm side} = 64$, which corresponds to a pixel size of
$\sim$0.85 deg$^2$. 

\begin{figure*}
\subfigure{\includegraphics[width=8.5cm]{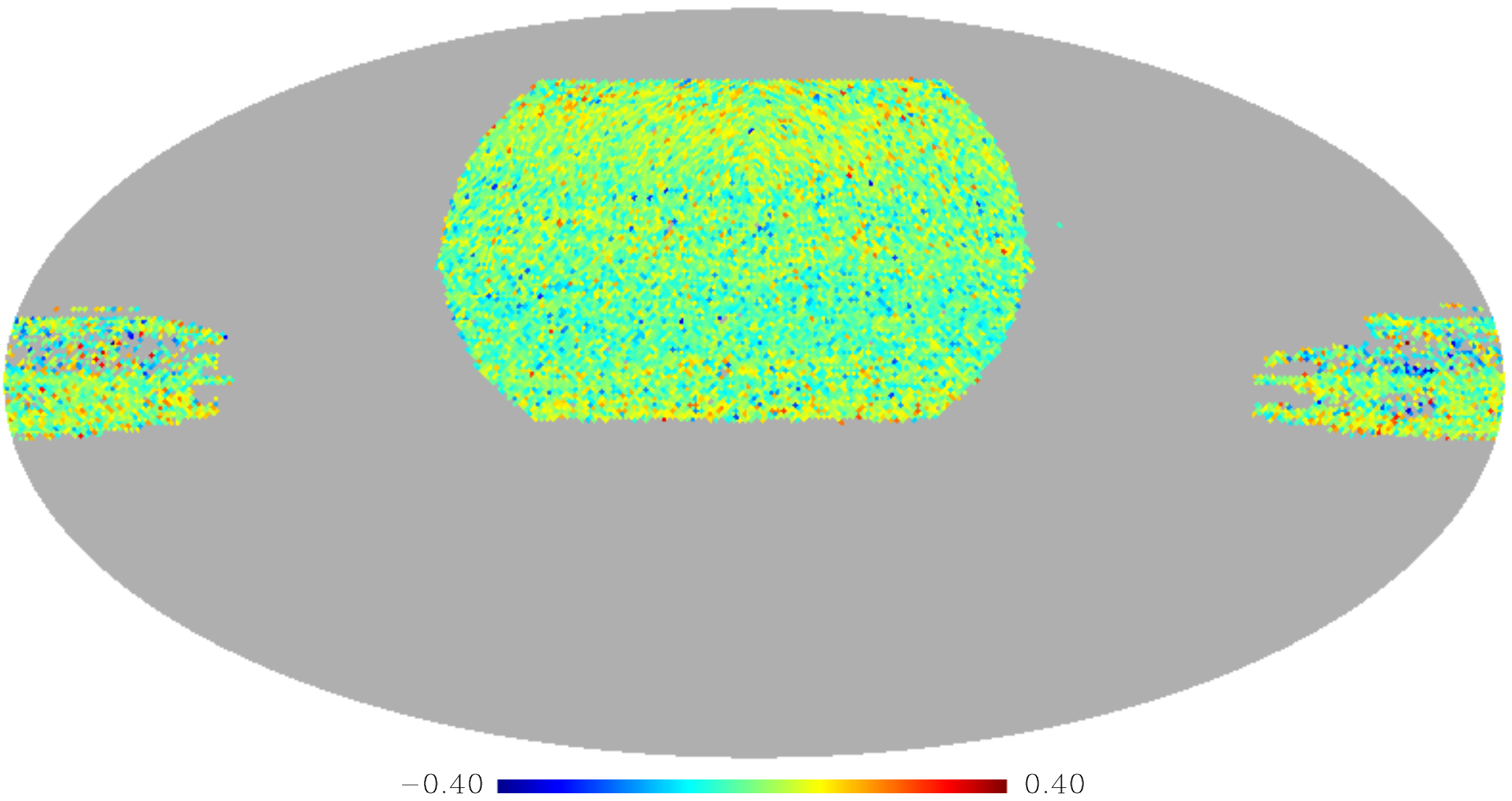}}\hspace{0.5em}
\subfigure{\includegraphics[width=8.5cm]{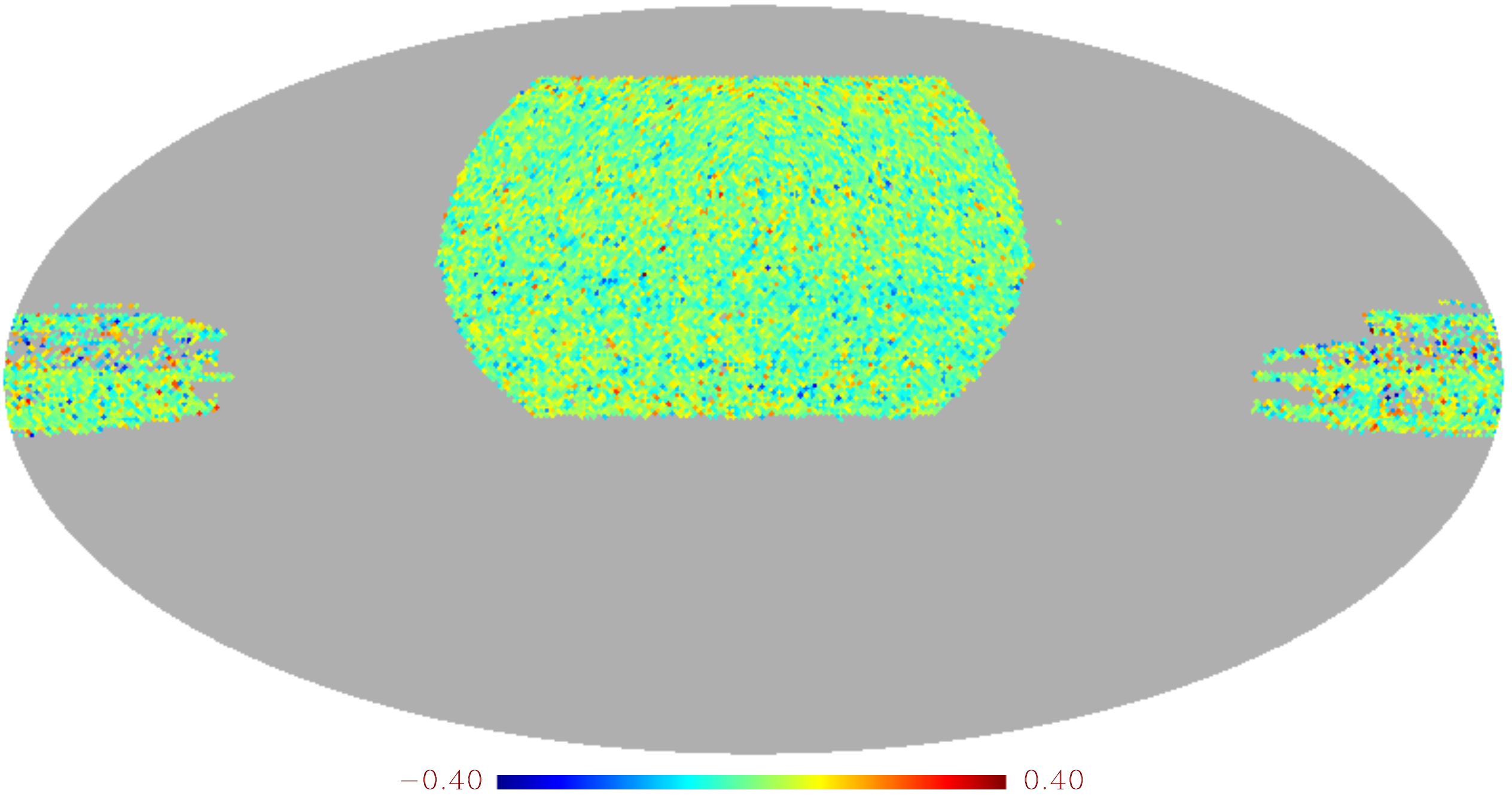}}\hspace{0.5em}
\caption{Maps of the $\gamma_1$ (\emph{left panel}) and $\gamma_2$
  (\emph{right panel}) shear components, constructed by simple
  averaging of the FIRST galaxy ellipticities within each pixel.}
\label{fig:first_maps}
\end{figure*}

Fig.~\ref{fig:first_maps} shows maps of the shear components,
$\gamma_1$ and $\gamma_2$, as reconstructed from the FIRST galaxy
ellipticities. These maps are dominated by random noise due to the
intrinsic dispersion in galaxy shapes and measurement errors. Visual
inspection of the figure does also suggest a weak dependence of
the $\gamma_1$ amplitudes on declination though some of this may
simply reflect the varying galaxy number density, which also shows some
dependence on declination. (Later in Section~\ref{sec:rr}, we perform a
suite of null-tests on our cross-power spectrum results, one of which
is to split the datasets according to declination. This test should
highlight any significant problems associated with any
declination-dependent systematic effects that persist in the FIRST
shear maps.)

\begin{figure} 
\centering
\includegraphics[width=8.5cm]{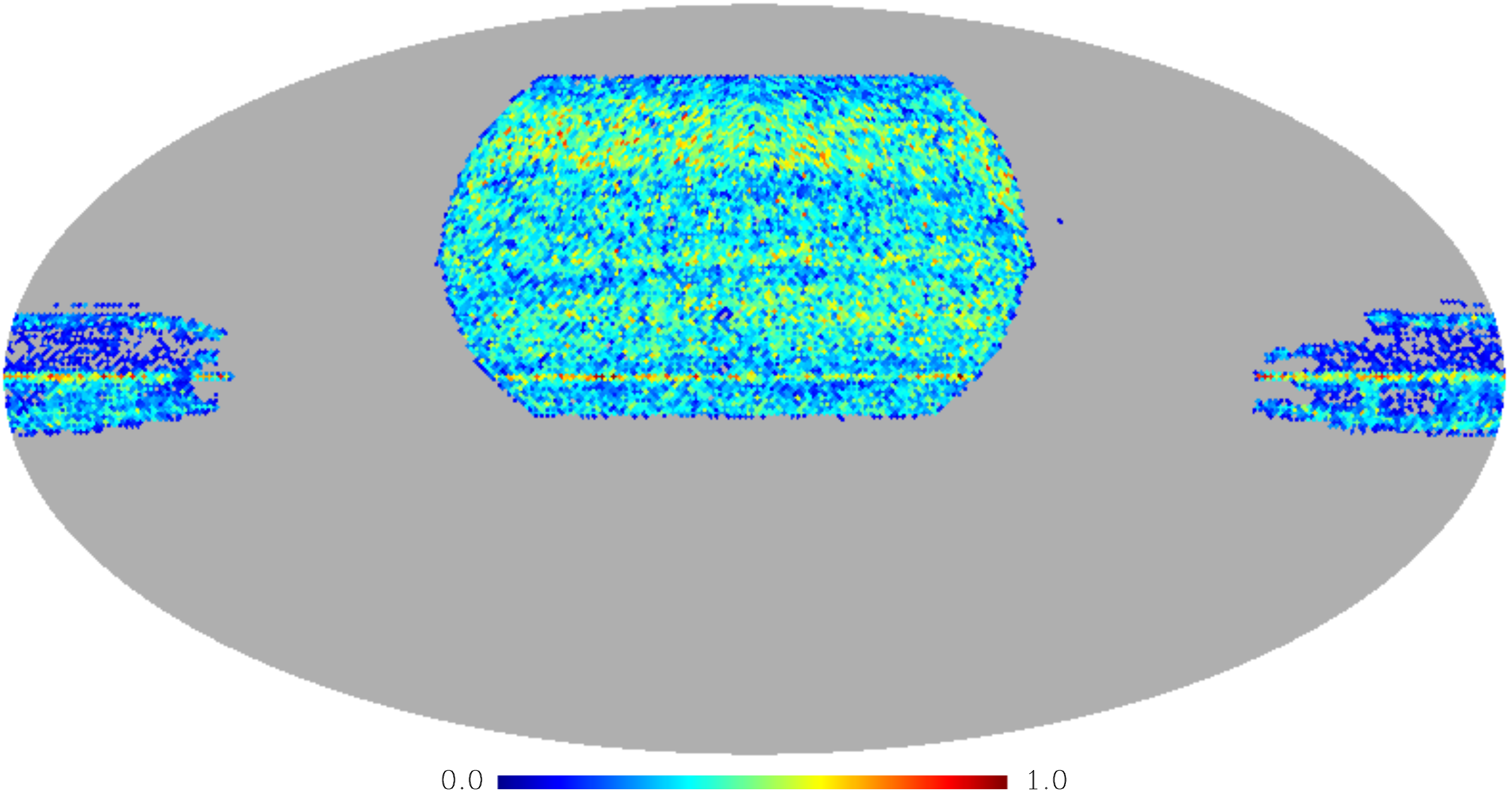} 
\caption{The galaxy number density  map used to weight the FIRST
  shear field (Fig.~\ref{fig:first_maps}) in the power spectrum
  analysis.}
\label{fig:first_mask}
\end{figure}

In addition to the shear maps, we construct a weight map (the $W(\Omega)$
of equation~\ref{eq:weighted_shear_field}) as simply the number of
galaxies that lie within each pixel. In the case of uniform
measurement errors on the ellipticities, setting $W(\Omega)$ to be
equal to the galaxy number density is equivalent to an
inverse-variance weighting scheme, which, for the noise-dominated
r\'egime in which we are working, is the optimal weighting scheme to
use. To limit our sensitivity to individual outliers in the
galaxy ellipticity distribution, we set $W(\Omega) = 0$ for pixels
that contain less than 5 galaxies. Fig.~\ref{fig:first_mask} shows the
weight map that results from this process. We use this map to
weight the FIRST shear data in the subsequent power spectrum analysis. 

\subsection{SDSS analysis}\label{sec:sdss_syst_and_maps}
\subsubsection{Shape measurements and initial source
  selection}\label{sec:sdss_src_select}
The SDSS catalogue provides shape information for each detected source
as ellipticity components derived from second moments,
\ba
\epsilon_1 = \frac{Q_{xx} - Q_{yy}}{Q_{xx} + Q_{yy}}, \\
\epsilon_2 = \frac{2Q_{xy}}{Q_{xx} + Q_{yy}},
\ea
where the adaptive moments are measured directly from the SDSS images
according to 
\be
Q_{ij} = \int d\vOmega I(\vOmega) w(\vOmega) \Omega_i \Omega_j.
\ee
Here, $I(\Omega)$ is the object intensity at position $\vOmega =
(\Omega_x, \Omega_y)$ and $\{i,j\}$ can take values $\{x,x\}$, $\{y,y\}$ or
$\{x,y\}$. Ideally, adaptive moments are measured using a radial weight
function, $w(\vOmega)$, which is iteratively adapted to the object's size. In
practice, the SDSS pipeline used a Gaussian weighting function with a
size matched to that of the object being fitted. Such an approach
has been shown to be nearly optimal \citep{bernstein2002} and is much
quicker. 
\begin{figure*}
\subfigure{\includegraphics[width=8.5cm]{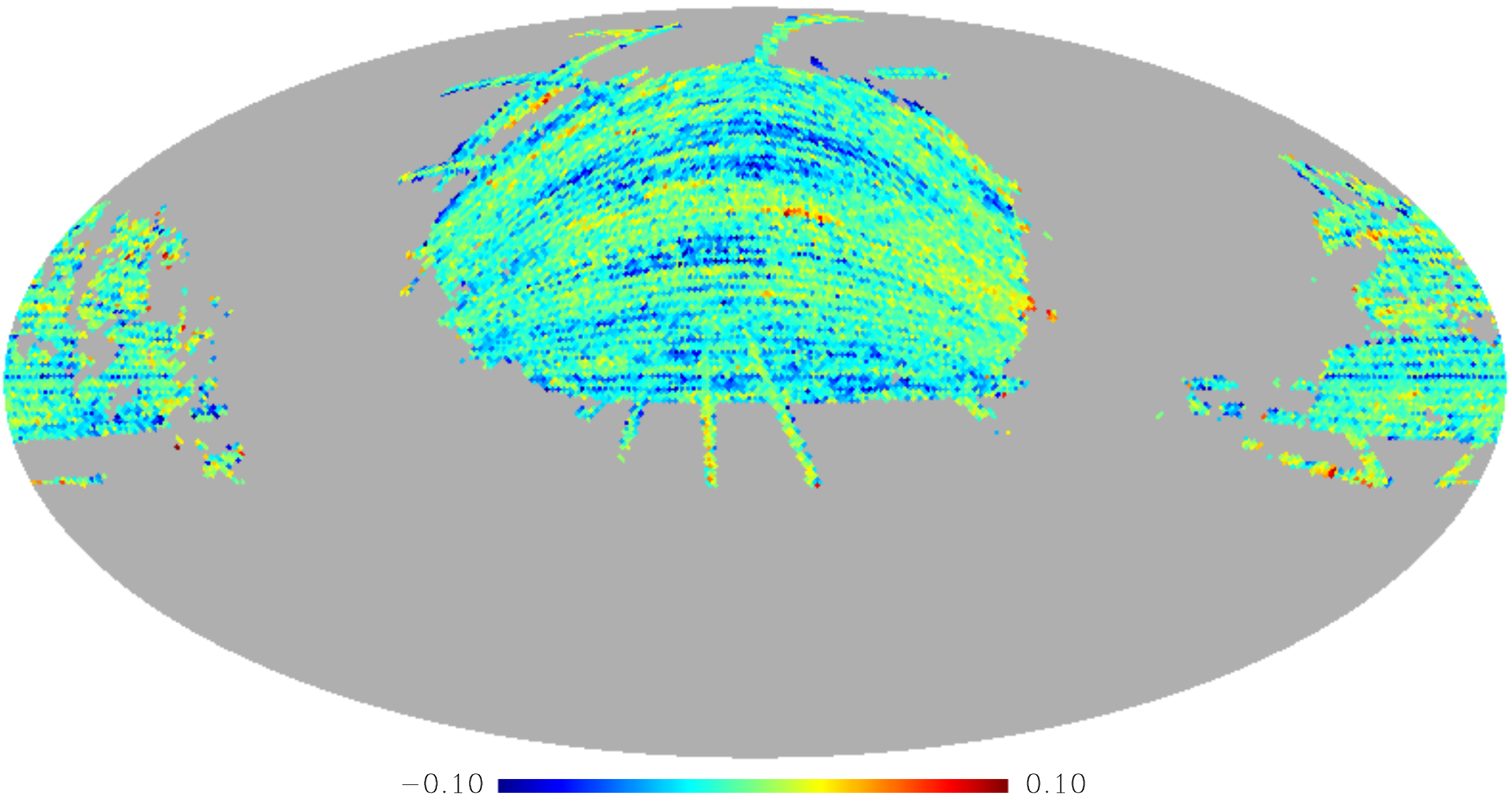}}\hspace{0.5em}
\subfigure{\includegraphics[width=8.5cm]{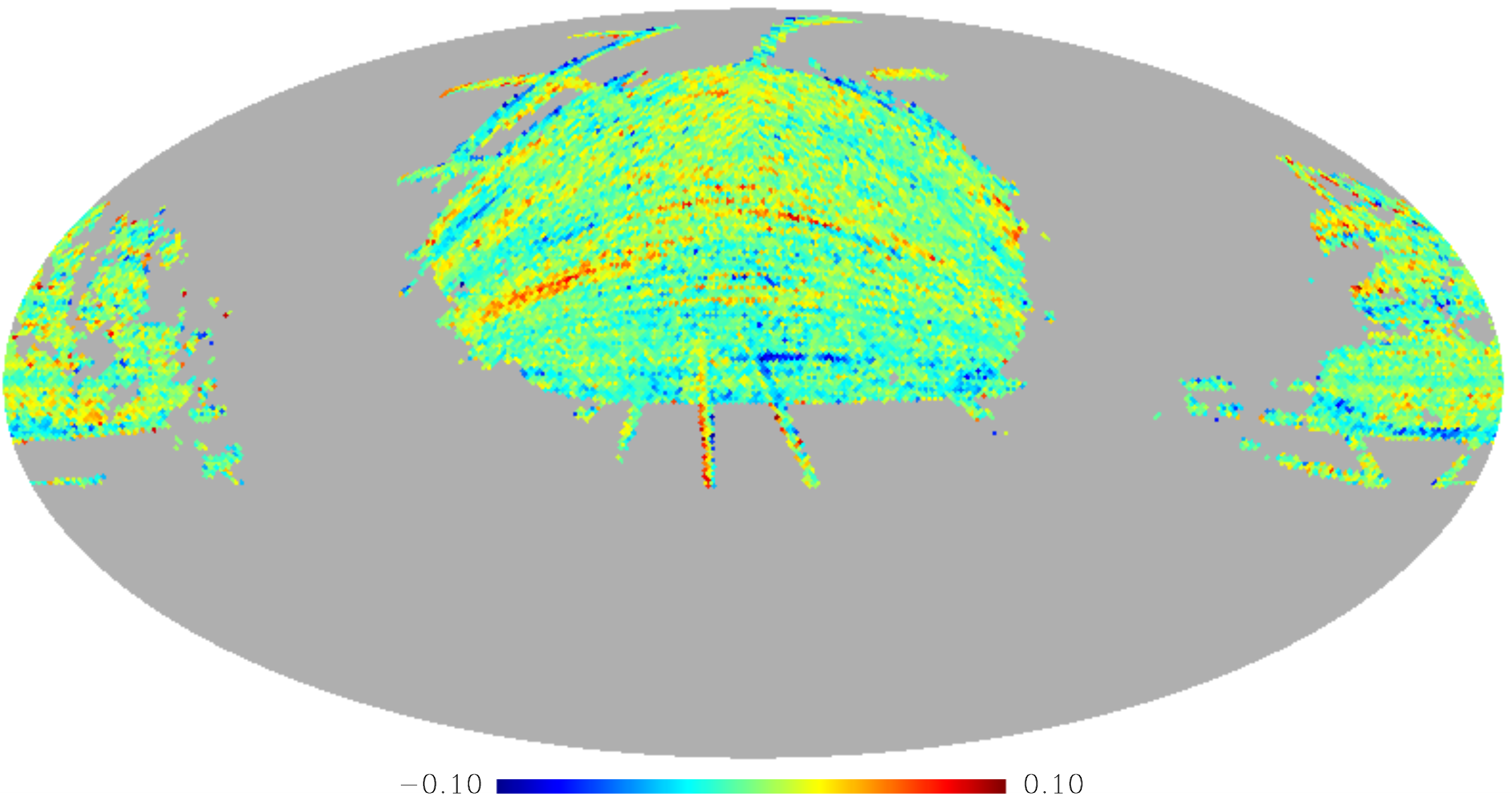}}\hspace{0.5em}
\caption{Maps of the $\gamma_1$ (\emph{left panel}) and $\gamma_2$
  (\emph{right panel}) shear components, constructed by simple
  averaging of the SDSS galaxy ellipticities within each pixel. These
  maps were constructed from the $\sim$25 million galaxies remaining
  in the SDSS shear catalogue immediately after the PSF correction and
  source selection steps described in Section~\ref{sec:sdss_src_select}
  were implemented. Significant large-scale systematic effects are
  clearly evident in these maps.}
\label{fig:sdss_maps1}
\end{figure*}
The SDSS catalogue's galaxy ellipticities need to be corrected for PSF
smearing and atmospheric seeing. To facilitate this, the
catalogue also provides moment information that describes the PSF
reconstructed at each galaxy's position. We follow the procedure
advocated by the SDSS team which corrects the catalogue ellipticities
for the effects of seeing and PSF anisotropy according to 
\ba
\epsilon_1^{corr}=(\epsilon_1^{meas} - R \epsilon_1^{psf})/(1-R) \nn
\epsilon_2^{corr}=(\epsilon_2^{meas} - R \epsilon_2^{psf})/(1-R),
\label{eq:sdss_psf_correction}
\ea
where $\epsilon_i^{meas}$ are the galaxy ellipticity components,
$\epsilon_i^{psf}$ are the PSF ellipticity components reconstructed at
each galaxy position and $R$ is a parameter formed from a combination
of the second and fourth-order moments of the galaxy and PSF
shapes. (For further details, and an explicit formula for $R$, see the
description on the SDSS DR10 catalogue
webpage\footnote{\url{http://www.sdss3.org/dr10/algorithms/classify.php}}.)
As mentioned in the SDSS catalogue description, the correction of
equation~(\ref{eq:sdss_psf_correction}) is not exact and results in a
small bias \citep{bernstein2002, hirata2003}. However, rather than
implementing a more sophisticated correction algorithm, we will mostly appeal to
the cross-correlation approach of Section~\ref{sec:science} to limit
our sensitivity to any residual biases that remain in the shear
catalogue after source selection and implementing the PSF correction
of equation~(\ref{eq:sdss_psf_correction}) (although see below for an
additional source selection cut that we have applied based on the
strength of the reconstructed PSF anisotropy, prior to construction of
our final shear maps).

To limit the impact of very poorly measured and/or noisy shape
estimates, we perform an initial source selection on the SDSS galaxy sample. To
select galaxies, we follow the works of \citet{bernstein2002} and
\citet{mandelbaum2008}. We thus exclude sources for which the modulus
of the corrected ellipticity
$\epsilon = (\epsilon_1^2 + \epsilon_2^2)^{1/2} > 4$ or if the uncertainty in
the corrected modulus, $\sigma_{\epsilon}>0.4$. We also only retain
sources that have an i-band extinction that is less than 0.2. In
addition, we retain only sources that have both a PSF and DeVaucouleur
r-band magnitude greater than 22, and both a PSF and DeVaucouleur i-band
magnitude greater than 21.6. Finally, to exclude very small galaxies,
we only include sources with $R_f>1/3$ where the resolution factor,
$R_f$ is given by, 
\be
R_f = 1 - \frac{Q_{xx}^{psf} + Q_{yy}^{psf}}{Q_{xx} + Q_{yy}}.
\ee 
With these selection criteria applied, $\sim$25 million SDSS
sources remain in the catalogue.

\subsubsection{Shear systematics and additional source
  selection}\label{sec:sdss_shear_maps}

\begin{figure*}
\subfigure{\includegraphics[width=8.5cm]{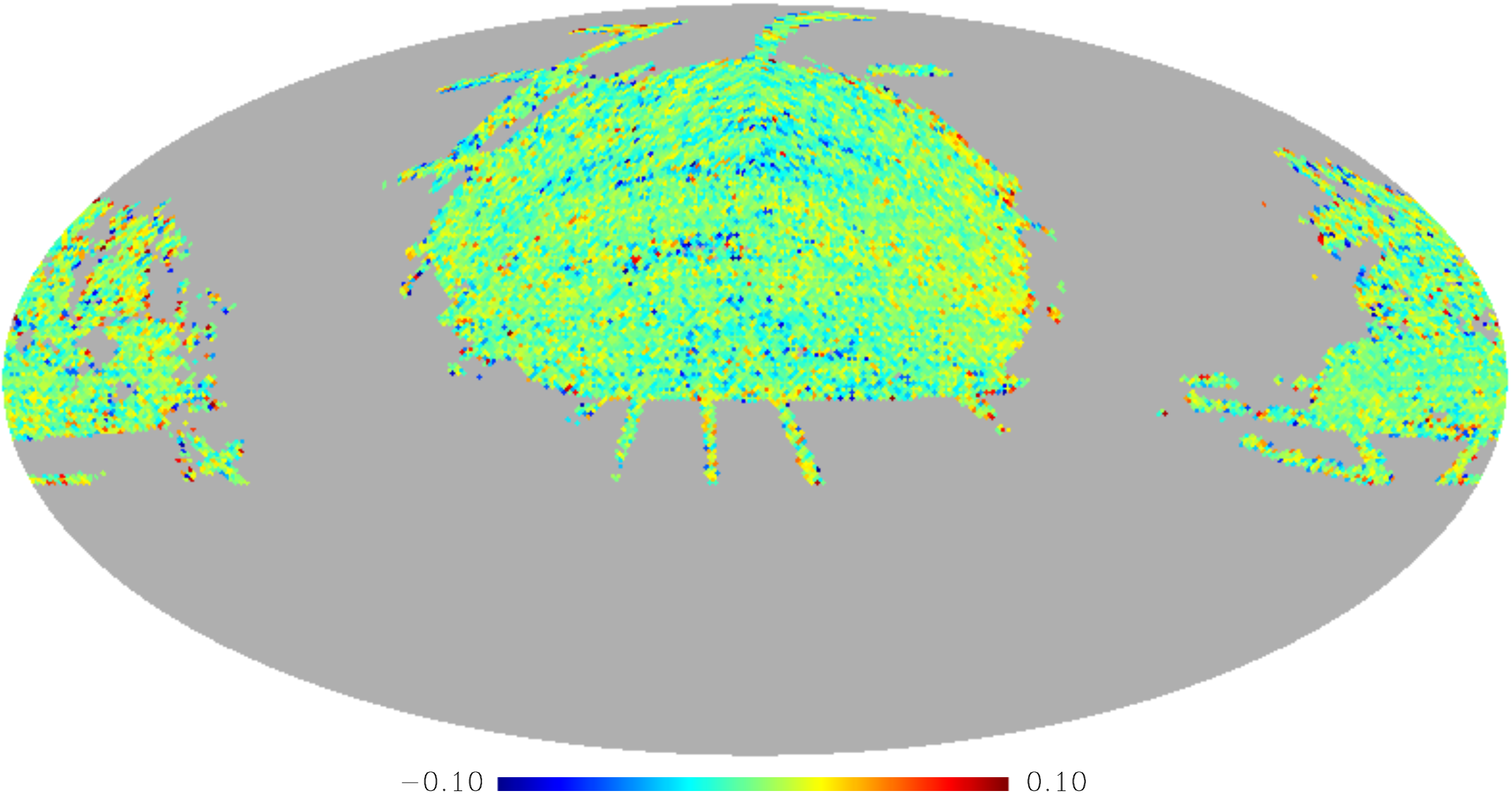}}\hspace{0.5em}
\subfigure{\includegraphics[width=8.5cm]{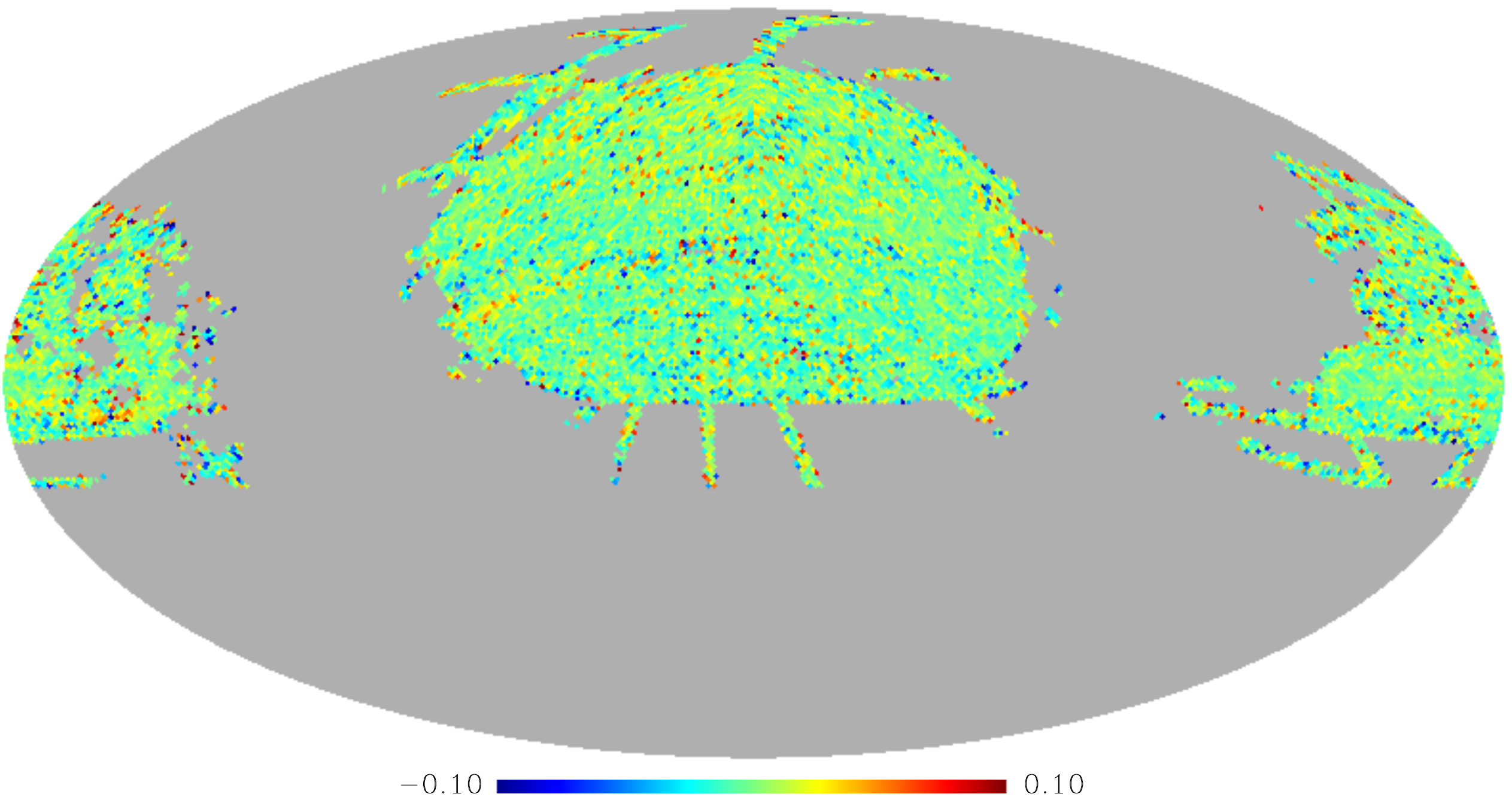}}\hspace{0.5em}
\caption{Maps of the $\gamma_1$ (\emph{left panel}) and $\gamma_2$
  (\emph{right panel}) shear components, constructed by simple
  averaging of the SDSS galaxy ellipticities within each pixel. These
  maps were constructed from the $\sim$9 million galaxies remaining in
  the SDSS shear catalogue immediately after the additional source
  selection based on the strength of the PSF, described in
  Section~\ref{sec:sdss_shear_maps}, was implemented. The 
  large-scale systematic effects apparent in Fig.~\ref{fig:sdss_maps1}
  have been substantially reduced with this additional cut on the galaxy
  catalogue.}
\label{fig:sdss_maps2}
\end{figure*}

Using the shape information of the remaining SDSS sources we construct
maps of the shear components $\gamma_1$ and $\gamma_2$ on a $N_{\rm
  side} = 64$ {\sevensize HEALPIX} grid using simple averaging, the
same as was used in the previous section for the FIRST maps. The
resulting maps are shown in Fig.~\ref{fig:sdss_maps1}.  Note
that, because of the large increase in galaxy numbers, the noise in
much lower in the SDSS maps than in the FIRST maps of
Fig.~\ref{fig:first_maps}. Visual examination of
Fig.~\ref{fig:sdss_maps1} shows that these initial SDSS shear maps are
clearly contaminated with large-scale spurious arcs. Comparing the
structure and morphology of these large-scale systematics with a map
of the reconstructed PSF anisotropy reveals a strong
correlation, indicating that the observed systematics are residual
biases that the approximate correction scheme of
equation~(\ref{eq:sdss_psf_correction}) has failed to account for.
 
In order to limit the impact of this residual bias, we have
implemented an additional cut on the galaxy catalogue based on the
strength of the PSF anisotropy at each galaxy position. We discard all
remaining SDSS sources for which the reconstructed PSF ellipticity
modulus is $\epsilon > 0.08$. (We have experimented with this
threshold value and have found $0.08$ to be the minimum
value which still suppresses the systematics to approximately the level of the
random noise.) Implementing this additional cut, we were left with $\sim$9
million SDSS sources. The SDSS shear maps constructed from this
revised catalogue are shown in Fig.~\ref{fig:sdss_maps2}. Visual
inspection of the new maps demonstrate a large reduction in the
amplitude of large-scale systematics, which now appear to be at or
below the level of the random noise. 

This additional cut on the galaxy catalogue increased the noise in the
SDSS shear maps by a factor $\sim\!\!\sqrt3$. However, the measured
SDSS auto shear power spectrum and the uncertainties associated with
the large-scale systematics were reduced by a factor of
$\sim\!\!10$. Although the FIRST-SDSS cross-power spectrum analysis
will be robust to this SDSS-specific systematic effect,
suppressing the PSF systematic to the level of the noise at this stage
in the analysis is still beneficial in terms of the error performance of the
cross-power spectrum analysis: without this additional selection based
on the PSF anisotropy amplitude, the uncertainties in the
cross-power spectrum increase by a factor of $\sim$\,$5$ (see
Section\,\ref{sec:rr} for more details).

\begin{figure} 
\centering
\includegraphics[width=8.5cm]{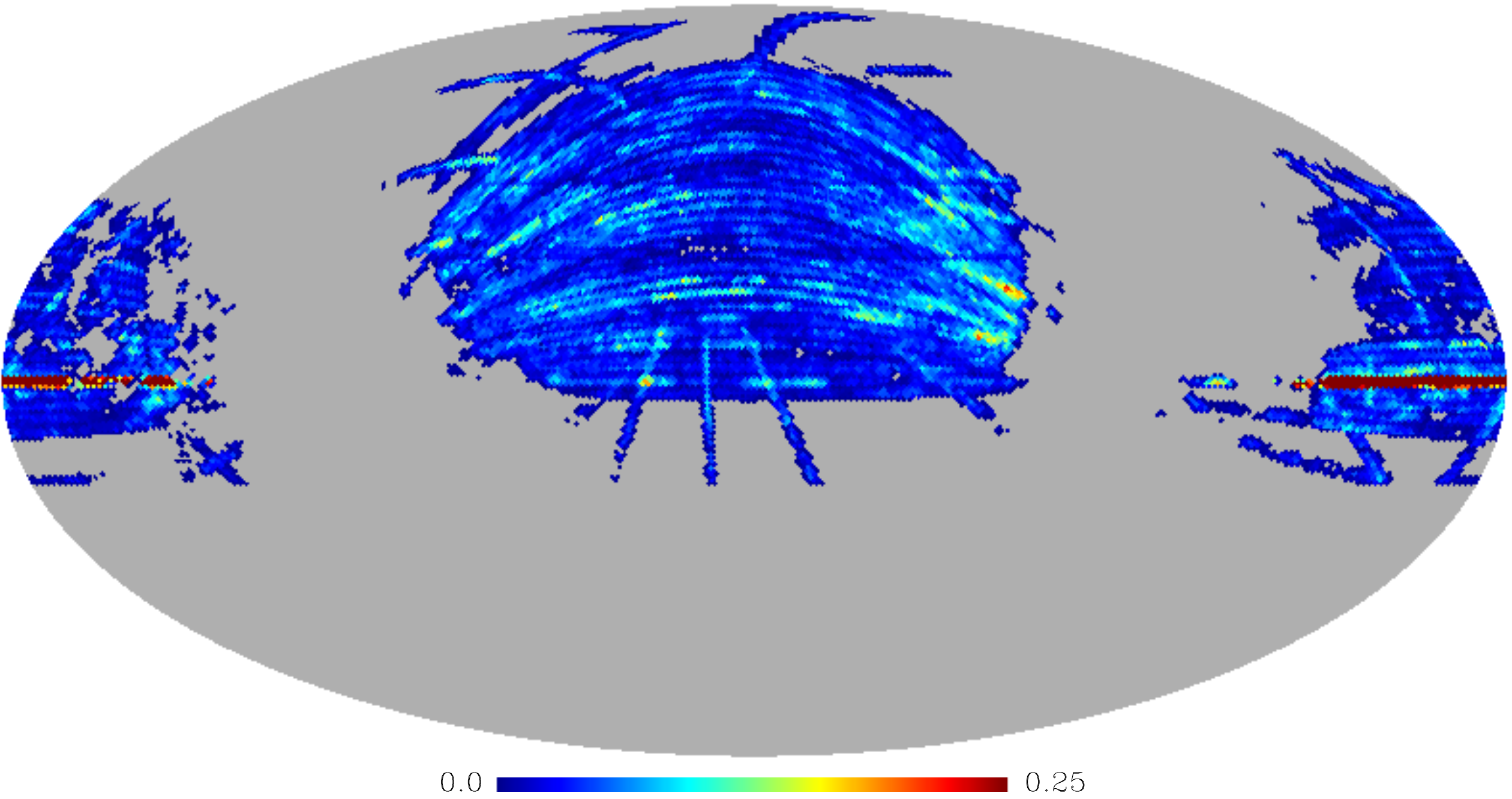} 
\caption{The galaxy number density map used to weight the SDSS
  shear field (Fig.~\ref{fig:sdss_maps2}) in the power spectrum
  analysis.}
\label{fig:sdss_mask}
\end{figure}

Finally, we generate a weight map for the SDSS survey using the same
method as was used in the previous section for the FIRST data. The
resulting weight map is shown in Fig.~\ref{fig:sdss_mask} and shows a
non-uniform distribution of SDSS source number density. In particular,
the very deep coverage of Stripe 82 is clearly visible as an excess of
galaxies near the Galactic anti-centre, at declinations $\delta
\sim$\,$0^{\circ}$.

\section{Simulations}\label{sec:sims}
In this section, we describe our approach to creating simulations of
the SDSS and FIRST shear datasets. Our simulations serve several
purposes. Firstly, the pseudo-$C_\ell$ power spectrum estimator
(equation~\ref{eq:bandpower_est}) includes a noise-subtraction step
which debiases a measurement of the shear power spectrum for the
effects of random shape noise (due to both intrinsic galaxy shape
noise and measurement errors). For this purpose, we estimate the noise
bias, $\lgl\widetilde{N_\ell}\rgl_{\rm MC}$ as the average
pseudo-$C_\ell$ power spectra measured from a suite of simulations
containing realisations of the random shape noise in each survey.

In addition, we will use simulations to estimate the uncertainties on
our power spectrum measurements. Given a set of simulations containing
both signal and noise components, we estimate the covariance matrix of
the power spectrum estimates as:
\be
\lgl \Delta \hat{P}_b \Delta \hat{P}_{b'} \rgl = \lgl (\hat{P}_b -\overline{P}_b)
(\hat{P}_{b'} - \overline{P}_{b'}) \rgl_{\rm MC},
\label{eq:bandpowers_covar}
\ee
where $\overline{P}_b$ denotes the average of each band power over all
simulations.

Finally, we can also use simulations to validate our power spectrum
analysis pipeline and to aid in the interpretation of null tests of
the results. 

\subsection{Signal simulations}
\label{sec:signal_sims}
We generate the signal component of our simulations based on a
$\Lambda$CDM cosmological model with parameter values taken from
\cite{planck2014}: $\Omega_m = 0.3175$, $\sigma_8 = 0.8347$, $H_0 =
67.1$ km s$^{-1}$ Mpc$^{-1}$, $\Omega_b = 0.0486$ and $n_s = 0.963$,
where $\Omega_m$ and $\Omega_b$ are the matter and baryon densities,
$\sigma_8$ is the clustering amplitude in 8 $h^{-1}$Mpc spheres, $H_0$
is the Hubble constant and $n_s$ is the spectral index of the
primordial perturbations. We additionally assume a flat Universe,
$\Omega_\Lambda = 1 - \Omega_m$.

\begin{figure} 
\centering
\includegraphics[width=8.5cm]{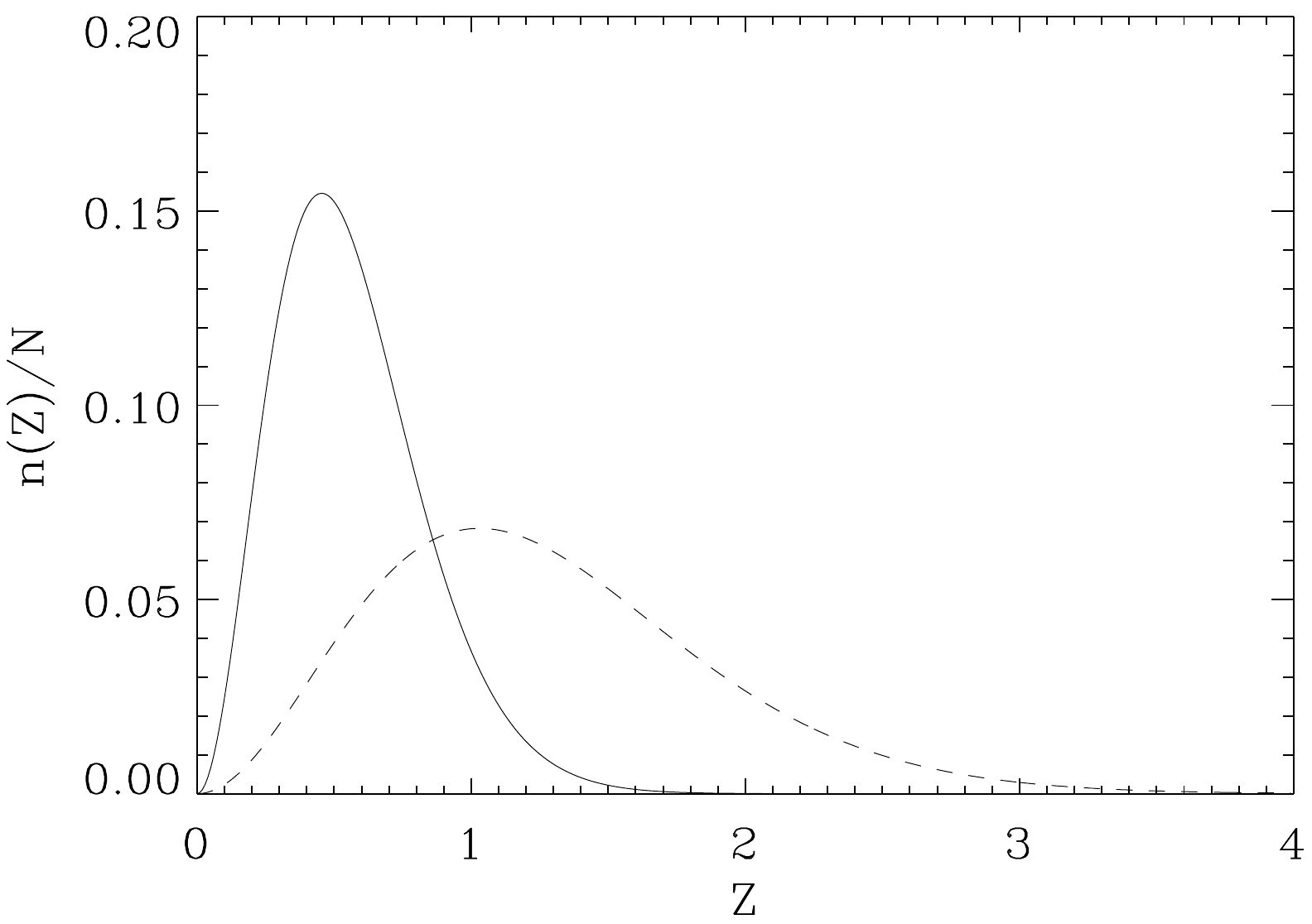} 
\caption{Normalised galaxy redshift distributions adopted for
  the SDSS population (solid line) and for the FIRST population
  (dashed line). We use these distributions to generate the model
  shear power spectra for our simulations.}
\label{fig:nofz}
\end{figure}

To generate the input model spectra, we also need to model the
redshift distributions of the two surveys. As mentioned earlier, using
the S$^3$ simulation~\citep{wilman2008} we have estimated the median
redshift of the FIRST survey to be $z_m^{\rm FIRST}=1.2$.  For SDSS,
we adopt a median redshift for the PHOTO catalogue of
$z_m^{\rm SDSS}=0.53$ from \citet{sypniewski2014}. This latter study
utilised a method for estimating photometric redshifts using boosted
decision trees and the known redshifts of the SDSS SPECTRO sample.
We then model both the SDSS and FIRST redshift distributions using the
parameterized form,
\be
n(z) = \beta \left( \frac{z^2}{z_*^3} \right) \exp \left[
  -\left( \frac{z}{z_*} \right)^{\beta} \right]~,
\ee
\label{eq:nofz}
where $\beta=1.5$, $z_* = z_m / 1.412$ and $z_m$ is the assumed median
redshift (0.53 for SDSS and 1.2 for FIRST). The resulting redshift
distributions are shown in Fig.~\ref{fig:nofz}. 

Using the above parameter values and $n(z)$ specifications, we generate
the three model shear power spectra (the auto power spectra for the
two surveys and the cross-power) according to
equation~(\ref{eq:clkk_theory}). For the 3-D matter power spectrum, we
use the \cite{bond1984} CDM transfer function. We also use the Halofit
formula \citep{smith2003} to predict the non-linear $P_\delta(k, r)$,
although for the scales of interest here, the non-linear contribution
is negligible.

We use these power spectra to generate correlated Gaussian 
realisations of the shear fields in the two surveys. To create the
correlated fields, at each multipole, $\ell$, we construct the $2
\times 2$ power spectrum matrix, $C_\ell^{\kappa(i)\kappa(j)}$ where
$\{i,j\}$ denote the two surveys. Taking the Cholesky decomposition of
this matrix at each multipole, $L_\ell^{ij}$, defined
by
\be
C_\ell^{\kappa(i)\kappa(j)} = \sum_k L_\ell^{ik} L_\ell^{jk},
\label{eq:cholesky_decomp}
\ee 
we generate random Gaussian realisations of the spin-2 spherical
harmonic coefficients of the two shear fields as
\ba
a^i_{\ell 0} &=& \sum_j L_\ell^{ij} G^j_{\ell 0}, \nn
a^i_{\lm} &=& \sqrt{\frac{1}{2}}\sum_j L_\ell^{ij} G^j_{\lm},
\label{eq:sph_harm_gen}
\ea
where $G^i_{\lm}$ is an array of unit norm complex Gaussian random
deviates. The resulting fields transformed to real space via a spin-2
transform will then be correctly correlated according to the input
auto and cross-power spectra. When generating the real-space shear
fields, we use the same spatial resolution as was used in constructing
the shear maps from the real data ({\sevensize HEALPIX} $N_{\rm side}
= 64$, corresponding to 0.85 deg$^2$ pixels). 

For each survey, we then create mock galaxy shape catalogues as
follows. For each object in the real catalogues, we replace its
measured ellipticity components with the components of the simulated
shear signal at the appropriate sky location in the {\sevensize
  HEALPIX}-generated maps. Note that we do not randomize the galaxy
positions. We choose to keep these fixed as we wish to
mimic the real data as closely as possible in the simulations. One
aspect of this is that we require the structure in the simulated
galaxy number density maps to be the same as that for the real data
(see~Figs.~\ref{fig:first_mask} and~\ref{fig:sdss_mask}). The observed
large-scale structure in the number density maps is clearly dominated by
observational selection effects rather than being due to intrinsic
galaxy clustering. It is therefore appropriate to use the exact same number
density fluctuations in the simulations in order to capture the impact
of these selection effects. Using the unaltered galaxy positions from
the real catalogues in all of the simulations ensures that this is the
case.

\subsection{Noise simulations}
\label{sec:noise_sims}
To create realisations of the random noise, we assign to each
simulated galaxy the measured ellipticity components of a real
galaxy, sampled at random from the real galaxy catalogues. 
This procedure should model the random noise properties of the data
(due to both the intrinsic dispersion in galaxy shapes and measurement
errors in the shape estimation step) and facilitate the estimation of
the noise bias term, $\lgl\widetilde{N}_\ell\rgl_{\rm MC}$. In
principle, our procedure may slightly over-estimate the noise since
the dispersion in the observed galaxy shapes will be enhanced slightly
due to any lensing signal present. However, we expect this effect to
have a negligible effect on our analysis -- the ellipticity variance
due to the combination of intrinsic shape dispersion and measurement
errors will greatly dominate over the lensing contribution for any
plausible cosmological signal. In any case, our cross-power spectrum
measurements are not affected by the noise bias term,
$\lgl\widetilde{N}_\ell\rgl_{\rm MC}$.

\subsection{Modelling small-scale systematic effects}
\label{sec:first_sys_sims}
\begin{figure} 
\centering
\includegraphics[width=8.5cm]{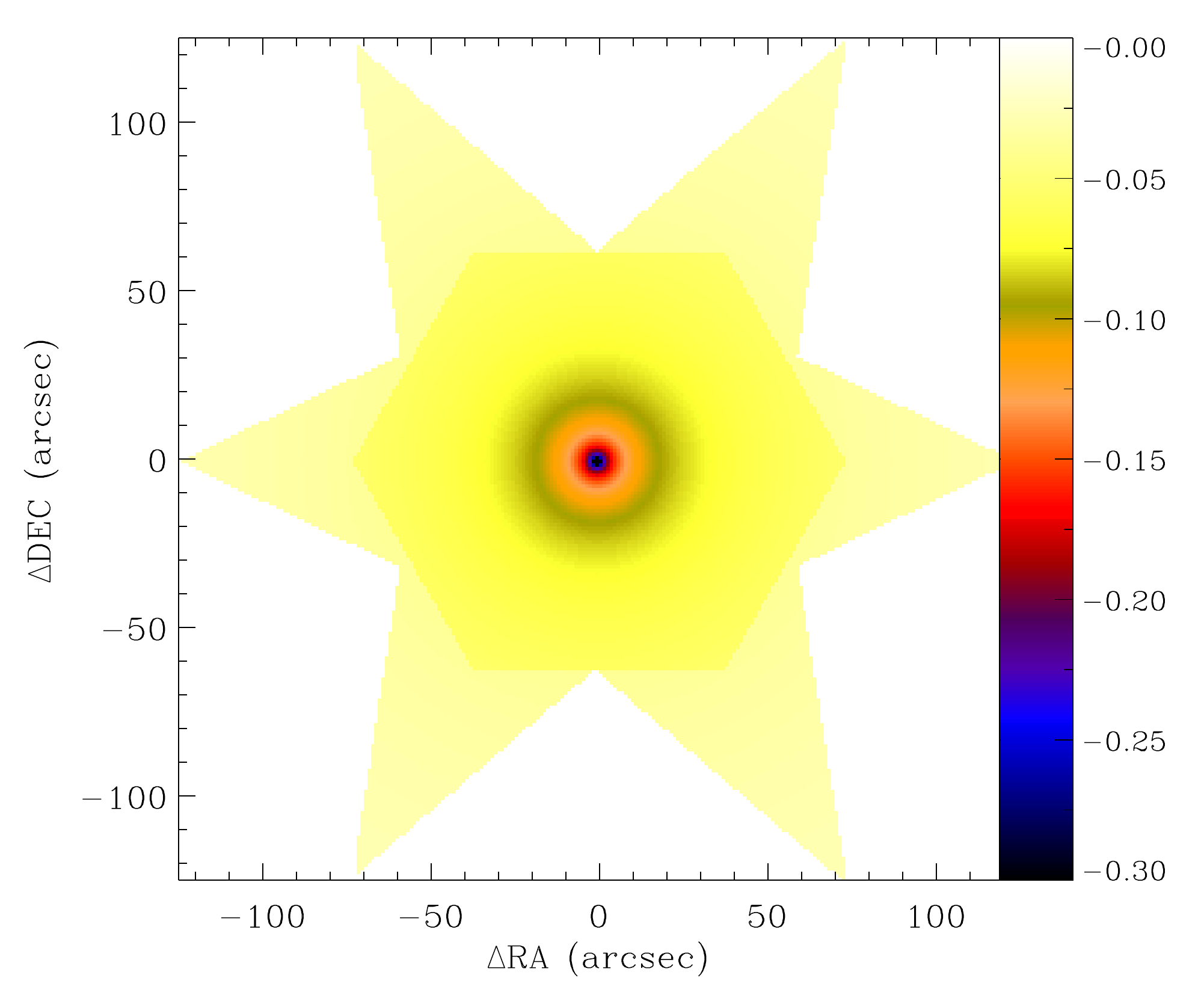} 
\caption{The $\gamma_t$ component of the contamination template used
  to model the FIRST residual beam systematic discussed in
  Section~\ref{sec:first_beam_sys}, and displayed in
  Fig.~\ref{fig:contdraddec}. The $\gamma_r$ component of the
  contamination was assumed to be zero.}
\label{fig:confade}
\end{figure}

In our FIRST simulations, we also include a model of the spurious
shear systematic, induced by the FIRST beam residuals, described in
Section~\ref{sec:first_beam_sys}, and illustrated in
Figs.~\ref{fig:contazim} \& \ref{fig:contdraddec}. To model this
effect, we first created a template of the observed spurious shear
signal. This template is shown in Fig.~\ref{fig:confade} and has been
normalised such that its azimuthally averaged tangential shear profile
matches that measured from the real FIRST survey, shown in
Fig.~\ref{fig:contazim}.

We then use this template to model the contaminating influence of each
source detected in the FIRST survey on all other FIRST sources. The
spurious signal induced on one source A, due to another source B, is
modelled as
\ba
\epsilon^{spur}_1 = \gamma_t(\Delta RA, \Delta \delta)
\cos(2\theta), \nn
\epsilon^{spur}_2 = \gamma_t(\Delta RA, \Delta \delta)
\sin(2\theta),
\label{eq:first_sys_sim}
\ea
where $\gamma^B_t(\Delta RA, \Delta \delta)$ is the tangential shear
extracted from the template at the position (in $\{\Delta RA, \Delta
\delta\}$-space) of source A relative to source B. In
equation~(\ref{eq:first_sys_sim}), $\theta$ describes the orientation
of the great circle connecting the two sources with respect to the
chosen reference axis of the $\gamma_{1/2}$ coordinate system.  

\begin{figure*}
\subfigure{\includegraphics[width=5.76cm]{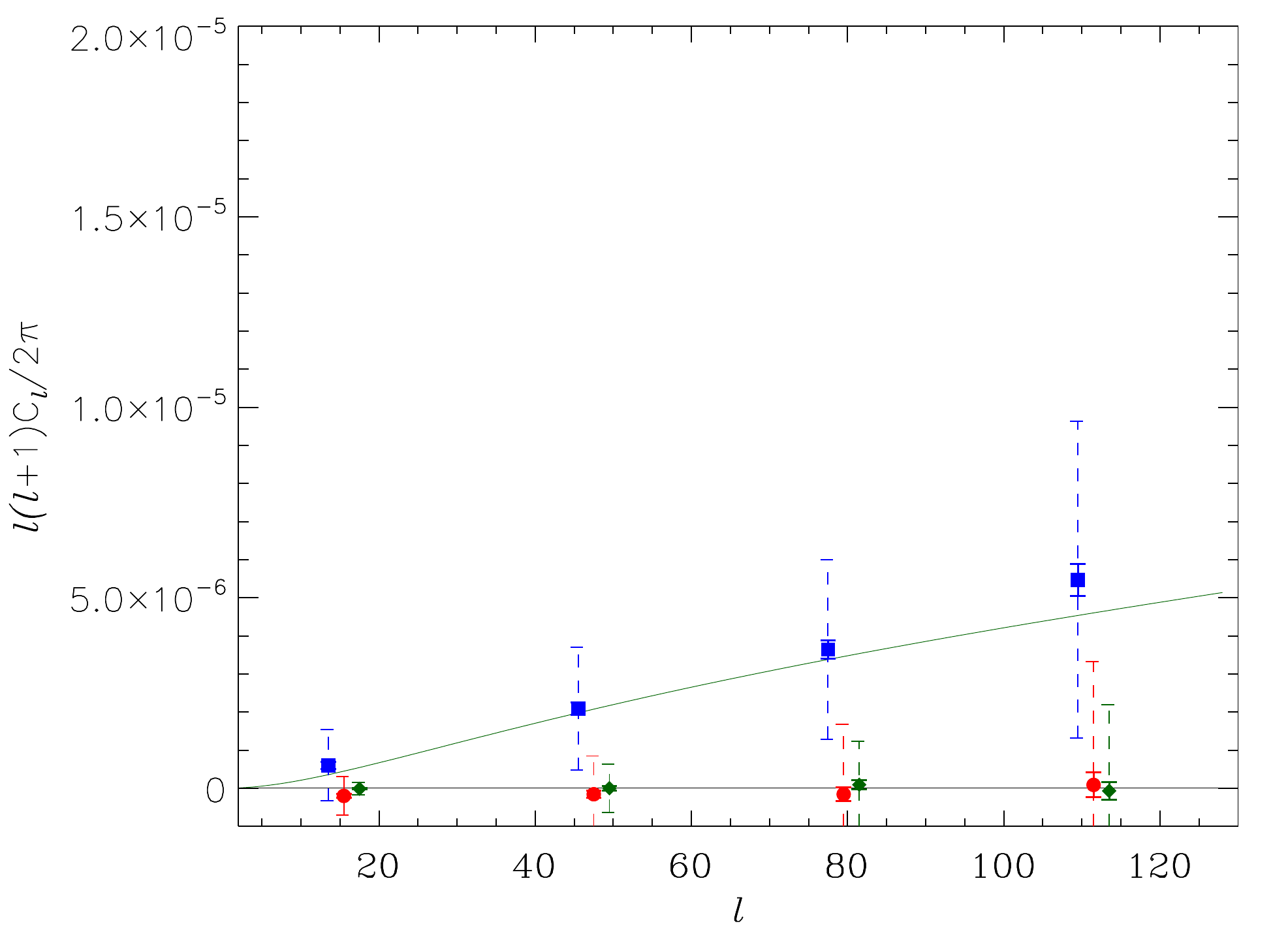}}\hspace{0.5em}
\subfigure{\includegraphics[width=5.76cm]{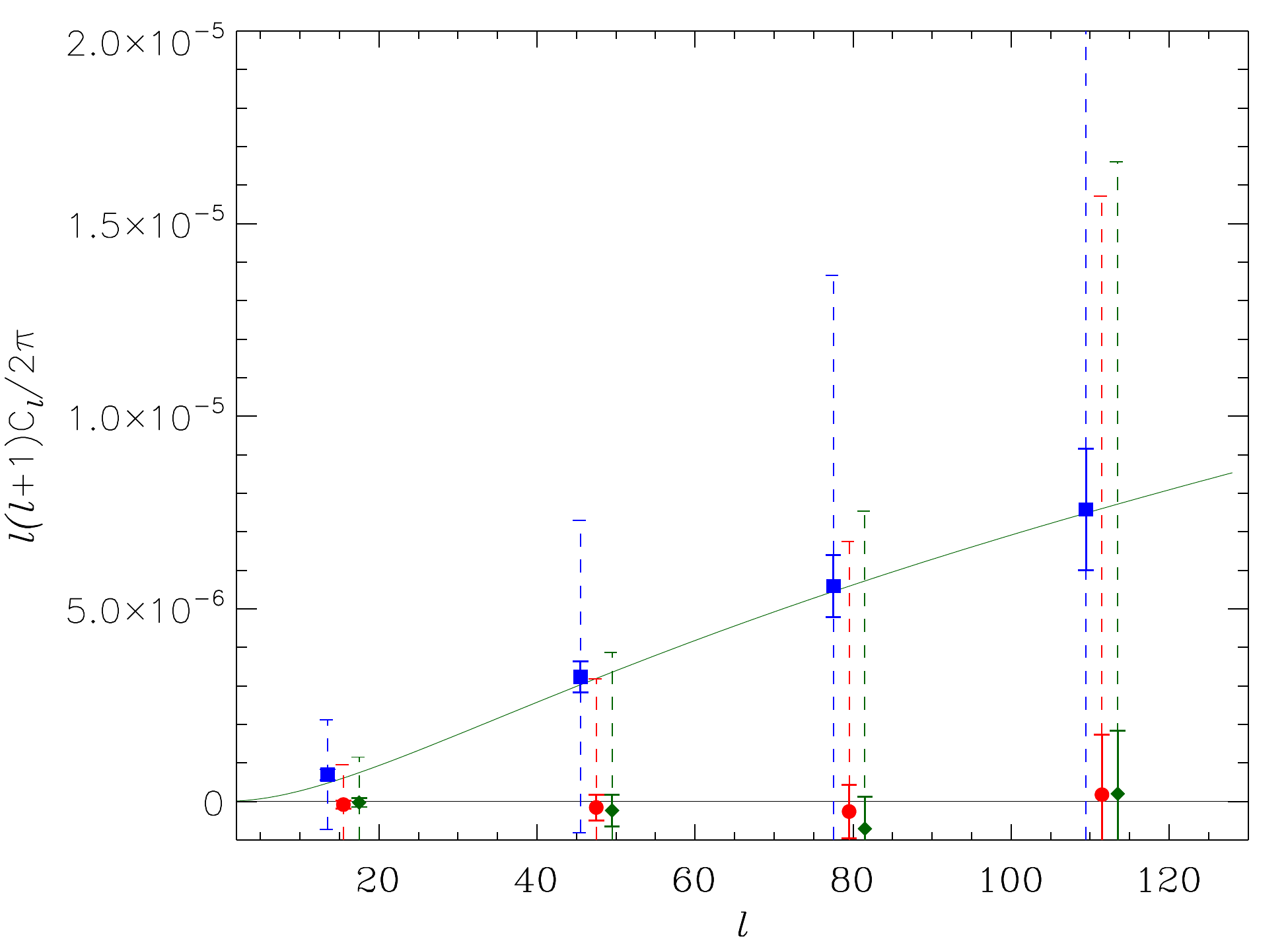}}\hspace{0.5em}
\subfigure{\includegraphics[width=5.76cm]{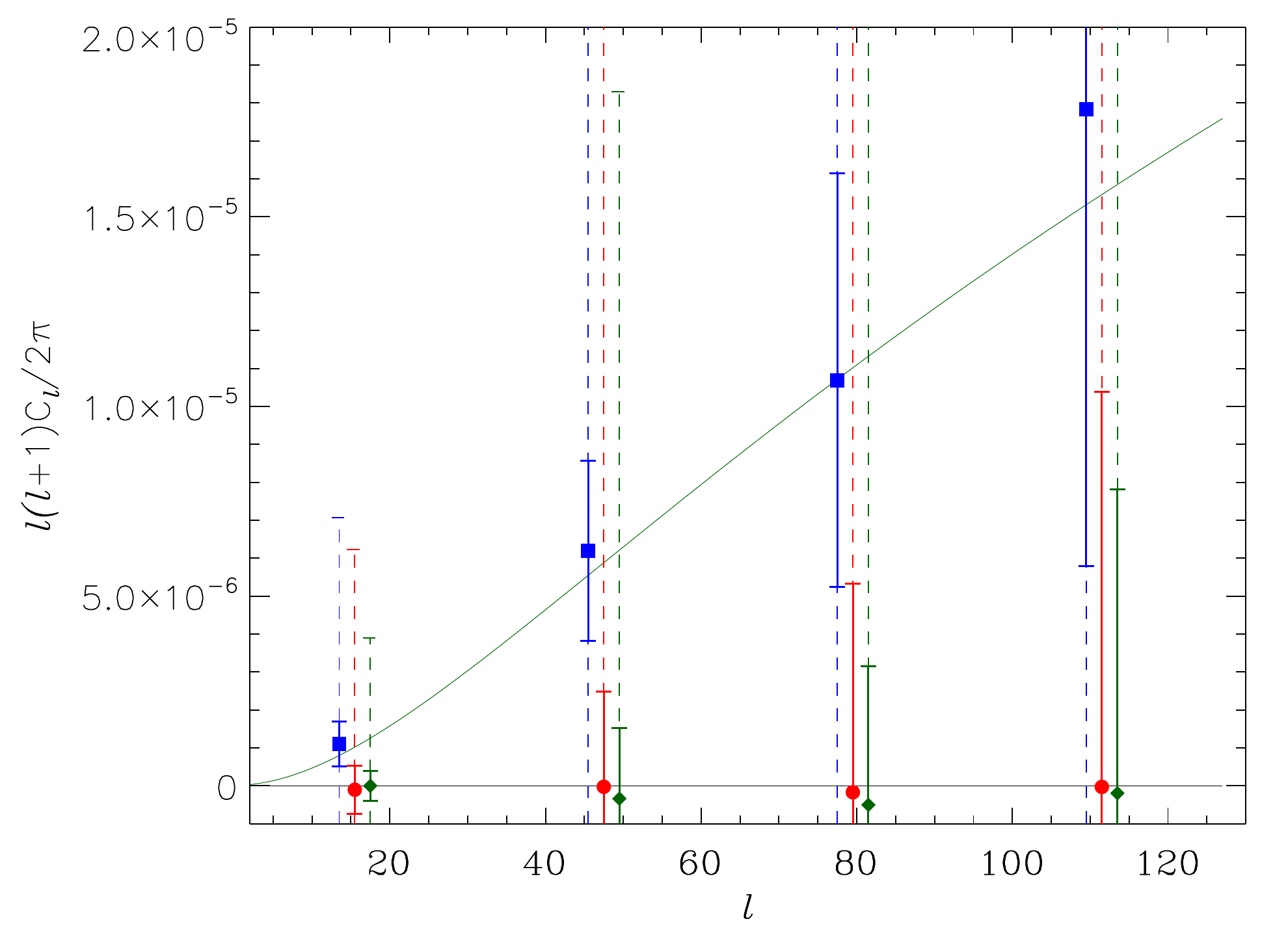}}\hspace{0.5em}
\caption{The recovered mean SDSS-SDSS (\emph{left}), SDSS-FIRST (\emph{centre}) and
  FIRST-FIRST (\emph{right}) $C_\ell^{\kappa\kappa}$ (blue squares), $C_\ell^{\beta\beta}$ (red
  circles) and $C_\ell^{\kappa\beta}$ (green diamonds) power spectra. The
  solid green curve shows the input $C_\ell^{\kappa\kappa}$ power spectra. The
  input $C_\ell^{\beta\beta}$ and $C_\ell^{\kappa\beta}$ power spectra
  were set to zero. Continuous and dashed line error bars show the
  error in the mean recovered values, and the uncertainty associated
  with a single realisation or measurement, respectively.}
\label{fig:avnoise}
\end{figure*}
\subsection{Power spectra from simulations}
\label{sec:sims_power_spectra}
We create 100 signal-only simulated datasets as described in
Section~\ref{sec:signal_sims}, each containing a known cosmic shear
signal, appropriately correlated between the FIRST and SDSS
realisations. Each realisation set also contains a noise-only
simulation (Section~\ref{sec:noise_sims}) which is uncorrelated
between datasets, and a model of the residual FIRST beam systematic
which is the same for each realisation
(Section~\ref{sec:first_sys_sims}). For each realisation, the sum of
all three of these components produces a full mock realisation of the
FIRST and SDSS shear catalogues. 

We process each simulated catalogue in exactly the same way as
we do for the real data. First, maps of the shear components are created
from the simulated FIRST and SDSS ellipticities using simple averaging
within $N_{\rm side} = 64$ {\sevensize HEALPIX} pixels. These maps are then
multiplied by the appropriate weight maps (Figs.~\ref{fig:first_mask}
\& ~\ref{fig:sdss_mask}) before they are passed through the power
spectrum estimation pipeline described in
Section~\ref{sec:pseudo_cl_theory}. 

Fig.~\ref{fig:avnoise} shows the mean recovered power spectra from the
100 simulations.  We estimate the three possible shear power spectra:
the SDSS auto power, the FIRST auto-power and the cross-power between
the two surveys. For each of these, in addition to the
$C_\ell^{\kappa\kappa}$ (or $E$-mode) power spectra, we have also
measured the $C_\ell^{\beta\beta}$ (or $B$-mode) power and the
cross-correlation, $C_\ell^{\kappa\beta}$ (or $EB$). We measure the
power spectra in four band powers ($1<\ell<32$; $33<\ell<64$;
$65<\ell<96$; $97<\ell<128$). For each band power we display two error
bars: the larger (dashed) errors show the uncertainty associated with
a single realisation or measurement (calculated as $\sigma_{\hat{P}_b}
= \lgl \hat{P}^2_b \rgl^{1/2}$ where $\lgl \hat{P}^2_b \rgl$ are the
diagonal elements of the covariance matrix,
equation~\ref{eq:bandpowers_covar}) while the smaller errors show the
error in the mean recovered band powers, $\sigma_{\overline{P}_b} =
\sigma_{\hat{P}_b} / \sqrt{N_{\rm MC}}$.

Inspecting the results, we see an unbiased recovery of the input
$C_\ell^{\kappa\kappa}$ signal in all cases while the recovered
$C_\ell^{\beta\beta}$ and $C_\ell^{\kappa\beta}$ are consistent with
zero, as required. The dashed error bars suggests that SDSS has the raw
precision to detect the cosmic shear signal on these scales, the
cross-power could potentially make a marginal detection while the
FIRST dataset is too noisy on its own to detect the signal. Note that,
on the relatively large scales of interest here, there is no
noticeable effect from the residual beam systematic that we have
included in our FIRST simulations. However, we have confirmed with
higher resolution simulations that this systematic does bias the
recovery of the FIRST auto-power spectrum, and increases the
uncertainties on the FIRST-SDSS cross-power spectrum, on smaller
angular scales (at higher $\ell$).

\subsection{Recovery in the presence of large-scale systematics}
\label{sec:sims_systematics}
The results of the previous section demonstrate that our power spectrum
analysis can recover a known input signal in the presence of both
random noise and the small-scale FIRST systematic. Those results also
show that our analysis correctly accounts for the effects of the
finite sky coverage and masking of the data.

However, visual inspection of Figs.~\ref{fig:first_maps} \&
\ref{fig:sdss_maps2} suggests that there are additional large-scale
systematics present in the data that are not modelled by our
simulations. Our approach to dealing with these systematics is a
central theme of this paper: by measuring the FIRST--SDSS cross-power
spectrum, these systematics should drop out of the analysis, provided
that the FIRST and SDSS systematic effects are uncorrelated. To
quantify the advantages of this approach, here we demonstrate the
unbiased recovery of a simulated cross-power spectrum signal in the
presence of either the real SDSS or the real FIRST large-scale
systematic effects.
 
To demonstrate unbiased recovery of the cross-power spectrum in the
presence of the SDSS systematics, we superimpose the SDSS
signal-only simulations onto the real SDSS map. Along with any signals
present in the real data (cosmic shear, noise and systematics), the
resulting coadded maps will now also contain a simulated signal that
is correlated with that present in the FIRST simulations. 
\begin{figure*}
\subfigure{\includegraphics[width=7cm]{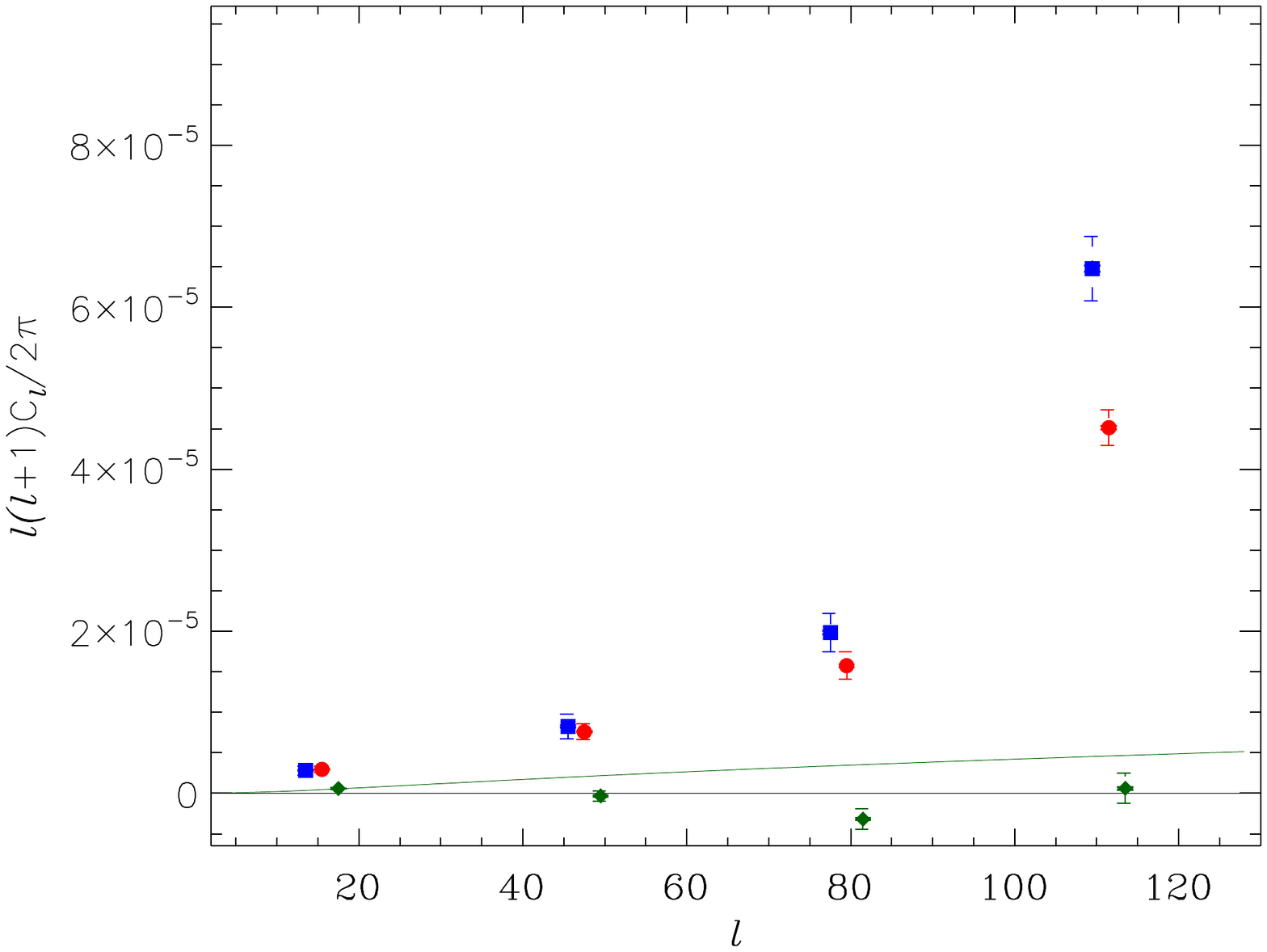}}\hspace{0.5em}
\subfigure{\includegraphics[width=7cm]{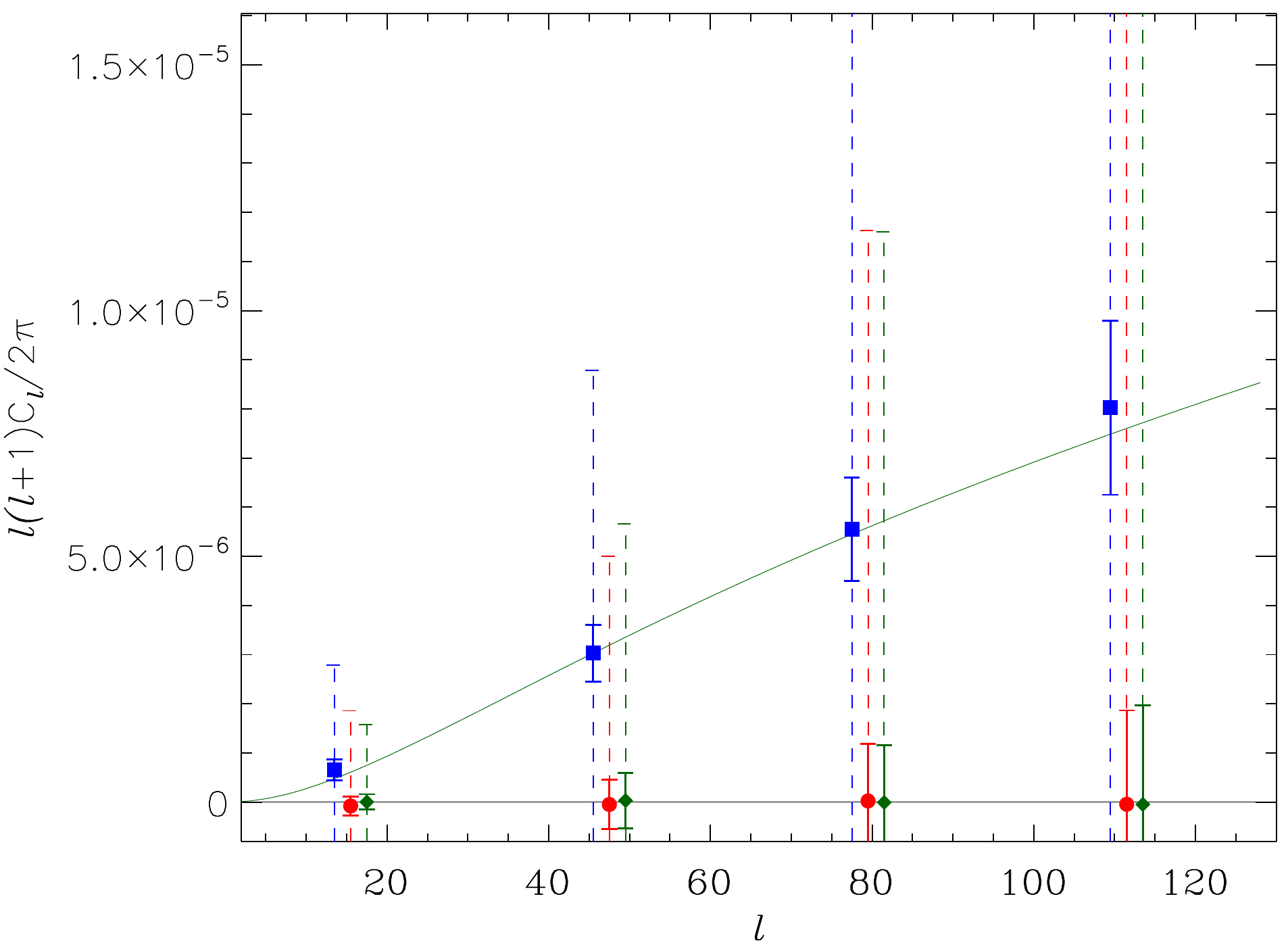}}\hspace{0.5em}
\caption{The recovered mean SDSS auto power spectra (\emph{left
    panel}) and SDSS-FIRST cross-power spectra (\emph{right panel})
  from 100 simulations in the presence of large-scale systematic effects in
  the real SDSS shear maps. For a definition of the symbols see
  Fig.~\ref{fig:avnoise}. Systematic effects in the SDSS data severely
  bias the recovery of the auto power spectrum, shown as the solid
  curve in the left hand panel. The cross-power spectrum recovery is
  unbiased in the presence of the same systematics, though the errors
  are amplified.}
\label{fig:Serr}
\end{figure*}
\begin{figure*}
\subfigure{\includegraphics[width=7cm]{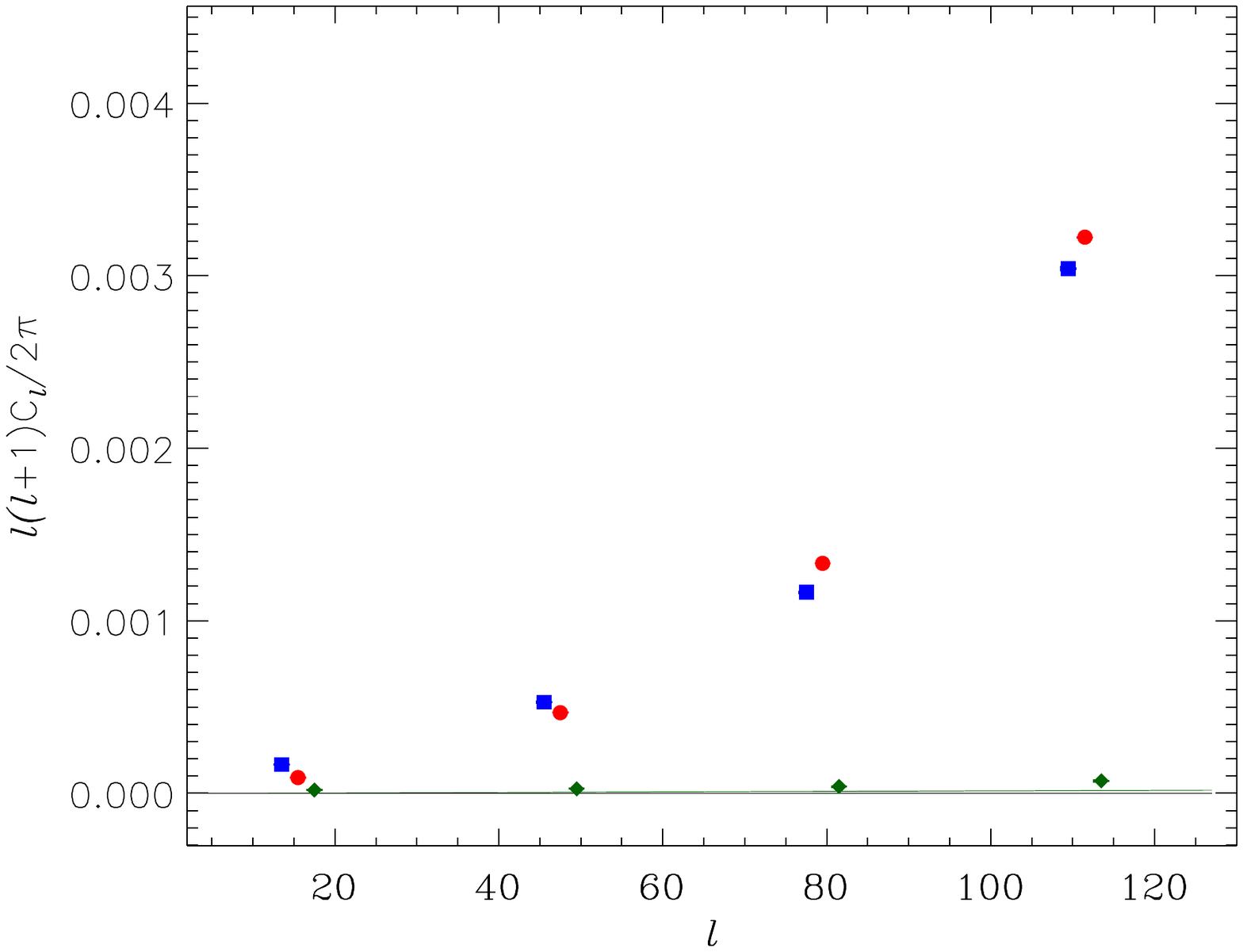}}\hspace{0.5em}
\subfigure{\includegraphics[width=7cm]{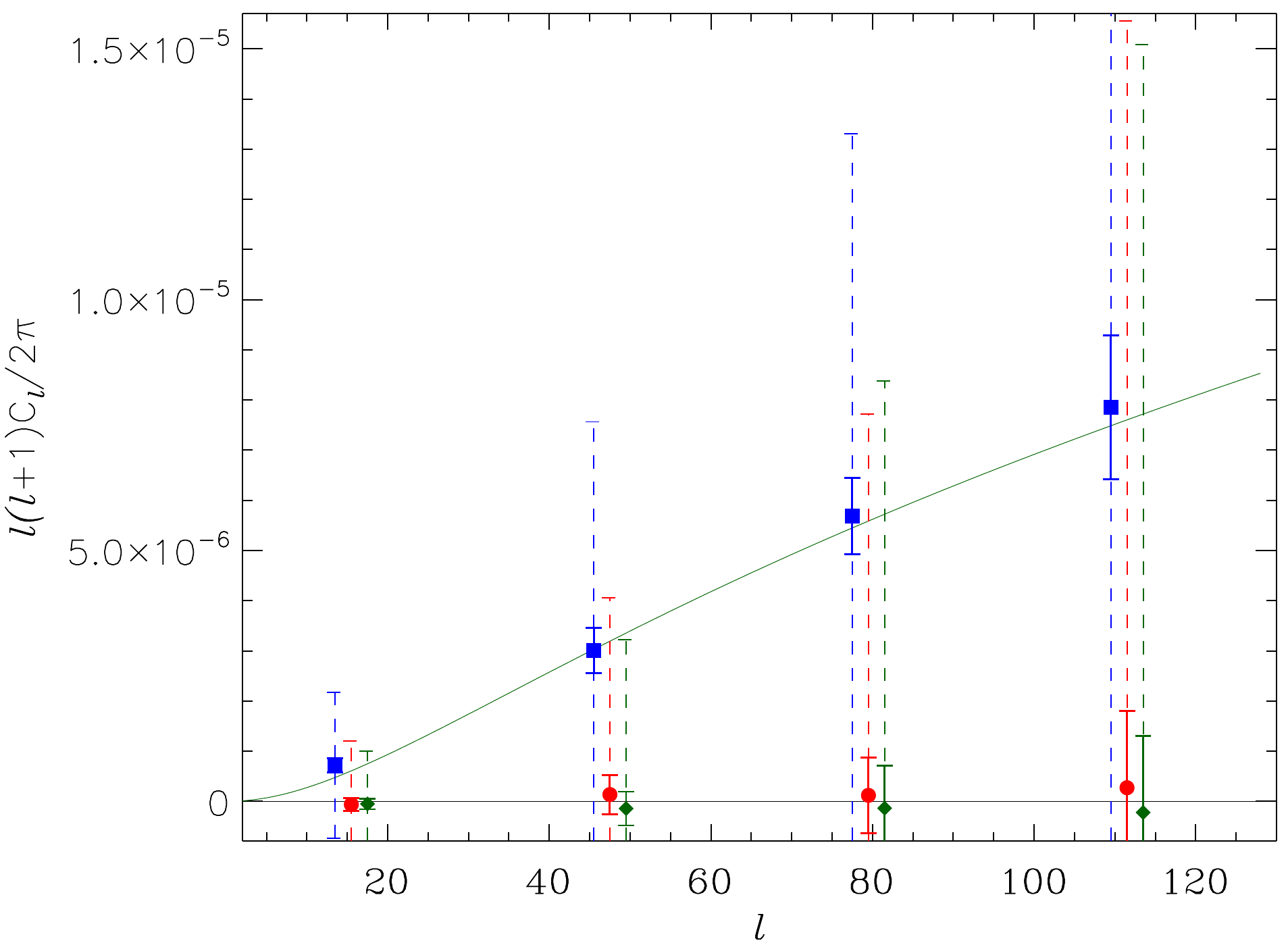}}\hspace{0.5em}
\caption{The recovered mean FIRST auto power spectra (\emph{left
    panel}) and SDSS-FIRST cross-power spectra (\emph{right panel})
  from 100 simulations in the presence of large-scale systematic effects in
  the real FIRST shear maps. For a definition of the symbols see
  Fig.~\ref{fig:avnoise}. Systematic effects in the FIRST data severely
  bias the recovery of the auto power spectrum, shown as the solid
  curve in the left hand panel. As in Fig.~\ref{fig:Serr}, the
  systematics contribute additional errrors into the cross-power
  spectrum measurement though it remains unbiased.} 
\label{fig:Ferr}
\end{figure*}
We have passed these coadded ``real + simulated'' maps through our
power spectrum estimation pipeline and the results are shown in
Fig.~\ref{fig:Serr}. These recovered power spectra have undergone
noise-bias subtraction and correction of the mask in exactly the same
way as before. Fig.~\ref{fig:Serr} shows that the recovery of the SDSS
auto power spectrum is very heavily biased by the presence of
unaccounted-for large-scale systematics in the data.\footnote{
The alternative explanation -- that the power observed in the left
hand panel of Fig.~\ref{fig:Serr} is of cosmological origin -- is
ruled out given current CMB and large-scale structure observations,
and in any case is incompatible with the FIRST-SDSS cross-power
spectrum that we measure later. The very large $C_\ell^{\beta\beta}$
signal observed is a further strong indicator that the observed
$C_\ell^{\kappa\kappa}$ signal in the SDSS auto power spectrum is due
to large-scale systematics.}
However, the right hand panel of Fig.~\ref{fig:Serr} shows that the
recovery of the SDSS-FIRST cross-power spectrum remains unbiased,
despite the presence of the SDSS systematics. Note that,
although it is unbiased, the cross-power spectrum is not completely
unaffected by the presence of the SDSS systematics: their presence
results in a significant increase in the error bars of the cross-power
spectrum recovery (compare the right-hand panel of Fig.~\ref{fig:Serr}
and the central panel of Fig.~\ref{fig:avnoise}), due to chance
correlations between the noise in the FIRST simulations and the SDSS
systematic effects. We will account for these additional contributions
to the errors in our final analysis.  

To assess the impact of the large-scale systematics in the FIRST shear
maps, we have performed an equivalent analysis to what is described
above but now we superimpose the FIRST signal-only simulations onto
the real FIRST maps and cross-correlate these with the SDSS
simulations. The results of this test are displayed in
Fig.~\ref{fig:Ferr} where we observe similar effects to what is
described above -- the FIRST auto-power spectrum is heavily biased by
large-scale systematics while the cross-power spectrum remains
unbiased.

\section{Real Data Measurements}\label{sec:rr}
In the previous section, we have demonstrated, using simulations, that
we can extract an unbiased cosmic shear signal by cross-correlating the two
datasets even in the presence of significant systematic effects in
both surveys, provided that the SDSS and FIRST systematics are not
correlated with one another. We now apply the analysis to the real
datasets.

\subsection{The SDSS-FIRST cross-power spectra}
To extract the cross-power spectrum from the real data, we apply the
exact same procedure as was followed for measuring the power spectrum
from the simulations in the previous section. Explicitly, we apply the
power spectrum estimator of equation~(\ref{eq:bandpower_est}) to
the real data maps, including both the noise debiasing step (based on the
mean noise bias, $\lgl N_{\rm MC} \rgl$, measured from noise only
simulations) and the correction for the survey geometries and masks. 

To estimate errors, we once again use the scatter in the MC
simulations (equation~\ref{eq:bandpowers_covar}). However, in addition
to the normal cosmic variance and noise terms, we now also wish to
include the enhanced uncertainties due to chance correlations between
random noise and large-scale systematics discussed in
Section~\ref{sec:sims_systematics}. To facilitate this, we generate
the \emph{total} mock cross-power spectrum for the each simulation
realisation as
\be
\hat{P}_b^{i} = \hat{P}_b({\rm sim}^i_F \times {\rm sim}^i_S) 
             + \hat{P}_b({\rm sim}^i_F \times {\rm real}_S) 
             + \hat{P}_b({\rm real}_F \times {\rm sim}^i_S), 
\label{eq:total_sim_power_spectra}
\ee
where $\hat{P}_b({\rm sim}^i_F \times {\rm sim}^i_S)$ is the
cross-power spectrum measured from the $i^{\rm th}$ simulation set
containing both signal and noise, $\hat{P}_b({\rm sim}^i_F \times {\rm
  real}_S)$ is the power spectrum measured by cross-correlating the
$i^{\rm th}$ FIRST simulation with the real SDSS map, and
$\hat{P}_b({\rm real}_F \times {\rm sim}^i_S)$ is the power spectrum
measured by cross-correlating the $i^{\rm th}$ SDSS simulation with
the real FIRST map. The covariance matrix of our measurements can then
be calculated from the scatter amongst the total mock power spectra of
equation~(\ref{eq:total_sim_power_spectra}), and will now include
uncertainites due to cosmic variance, random noise and the enhanced
uncertainties due to the systematic effects.

\begin{figure} 
\centering
\includegraphics[width=8.5cm]{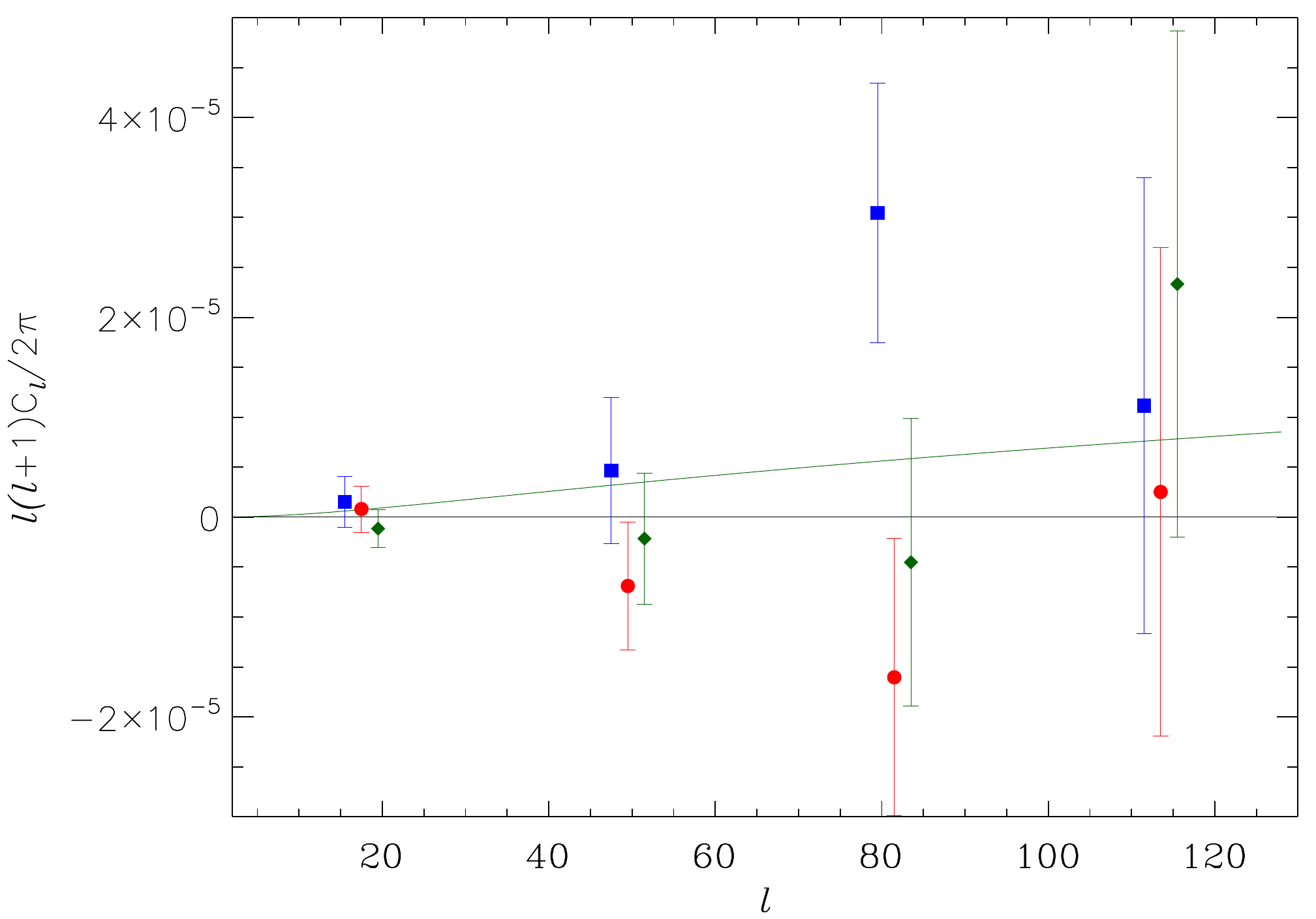} 
\caption{The cosmic shear cross-power spectrum measured from the SDSS
  and FIRST datasets. The blue squares show the measured
  $E$-mode signal ($C_\ell^{\kappa\kappa}$), the red circles show the
  measured $B$-mode ($C_\ell^{\beta\beta}$) and the green diamonds
  show the measured $EB$ correlation $C_\ell^{\kappa\beta}$. The green
  solid curve is the expected $C_\ell^{\kappa\kappa}$ power spectrum
  for two populations with median redshifts of 0.53 and 1.2
  respectively.}
\label{fig:res}
\end{figure}

The cross-spectrum measured from the real FIRST and SDSS datasets is
shown in Fig.~\ref{fig:res}. Comparing the real data measurement to
the distribution of measured power spectra from simulations that
include only noise and systematics, we find our measurement equates to
a marginal detection of the $C_\ell^{\kappa\kappa}$ power spectrum at
the 99\% confidence level (a ``$\sim\!\!2.7\sigma$ detection''). The
measured signal is in agreement (within $\sim\!\!1\sigma$) with the
model theory power spectrum where the median redshifts for the SDSS
and FIRST populations are assumed to be 0.53 and 1.2 respectively. The
$B$-mode power spectrum ($C_\ell^{\beta\beta}$) and the $EB$
cross-correlation ($C_\ell^{\kappa\beta}$) are both consistent with
zero (within $\sim\!\!1\sigma$).

\begin{figure*}
\centering \includegraphics[width=7in,height=3in]{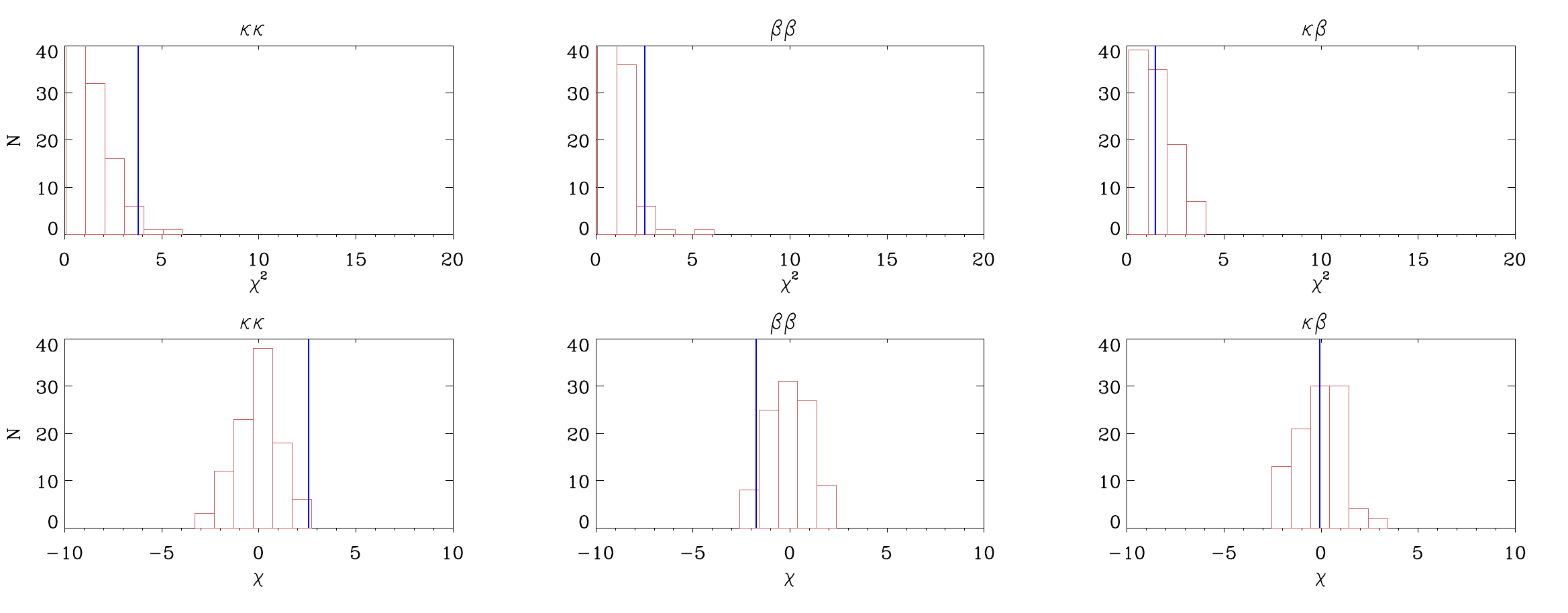}
\caption{Histograms showing the $\chi^2$ values
  (equation~\ref{eq:chi2}; \emph{upper panels}) and
  $\chi$ values (equation~\ref{eq:chi}; \emph{lower panels}) for the
  $C_\ell^{\kappa\kappa}$ (\emph{left}), $C_\ell^{\beta\beta}$
  (\emph{centre}) and $C_\ell^{\kappa\beta}$
  (\emph{right}) power spectra as measured from the
  simulations. Over-plotted as the vertical blue line is the
  equivalent value for the real data measurements.}
\label{fig:lcdm_chi2_tests}
\end{figure*}

To further assess the significance of our measurement and the degree to
which it is consistent with the expected signal in the concordance
cosmological model, we have calculated a $\chi^2$ statistic from our
suite of total simulated power spectra
(equation~\ref{eq:total_sim_power_spectra}) as:
\be
\chi^2 = \sum_b (\hat{P}_b - P_b^{\rm th})^2 / \sigma^2_{\hat{P}_b},
\label{eq:chi2}
\ee
where $P_b^{\rm th}$ is the expected value of the band power in the
concordance model. We then compare the $\chi^2$ value measured from
the real data to the distribution of values from the simulated data
sets. We have calculated this statistic for the
three power spectra ($C_\ell^{\kappa\kappa}$, $C_\ell^{\beta\beta}$ and
$C_\ell^{\kappa\beta}$) separately and note that, for the latter two spectra,
$P_b^{\rm th} = 0$. The results of this test are shown in the top
panels of Fig.~\ref{fig:lcdm_chi2_tests}. We see that our measurement from the real
data is consistent with the simulation distribution, and hence with
the expectation based on the concordance cosmology. 

The $\chi^2$ statistic tests for a particular type of
discrepancy between model and data. However, it does not test for all
possible deviations. In particular, it it insensitive to the sign of
$\hat{P}_b - P_b^{\rm th}$. We have therefore also performed a consistency
test using the following statistic (which we loosely call $\chi$):
\be
\chi = \sum_b (\hat{P}_b - P_b^{\rm th}) / \sigma_{\hat{P}_b}.
\label{eq:chi}
\ee
The results from this test are shown in the lower panels of
Fig.~\ref{fig:lcdm_chi2_tests} and once again, reveal no
inconsistencies between our measurements and the concordance
cosmological model. 

\subsection{Null tests}
\label{sec:null_tests}
We further assess the credibility of the data and results through a
set of measurements designed to reveal any residual systematics that,
for whatever reason, might not be mitigated using our
cross-power spectrum approach. For each of these null tests, we have
processed each of our simulated datasets in exactly the same way in
order to assign appropriate error bars to the null test
measurements. The tests that we have performed are the following:
\begin{enumerate}
\item \emph{North-South}: We split both the SDSS and FIRST data into North and South
  samples, with roughly the same number of sources in each sample. We then
  measure the power spectra for the two samples separately. Finally we
  subtract the signal between the two sets of spectra.
\item \emph{East-West}: This is the same as test (i) but splitting the
  data into East and West components.
\item \emph{FIRST random}: We randomly split the FIRST data into two subsets with
  the same number of galaxies. For each subset, a set of $\gamma_1$ and $\gamma_2$
  maps is then created. We take the difference of the shear maps for
  the two FIRST subsets and then estimate the cross-power spectrum of
  the resulting residual map with the SDSS shear map.
\item \emph{SDSS random}: The same as test (iii) but randomly
  splitting the SDSS data.
\item \emph{FIRST PSF}: This test is designed to test whether the measured signal is an
  artefact of the FIRST PSF. In addition to deconvolved shape
  estimates, the FIRST catalogue also includes estimates of source
  shapes before PSF deconvolution. We replace our FIRST
  galaxy ellipticity catalogue with an ellipticity catalogue based on
  these uncorrected shape measurements for only those sources that are
  flagged as point sources (unresolved) in FIRST. The shear
  maps constructed from this catalogue should then trace the FIRST PSF
  accurately. We then estimate the cross-power spectrum of the FIRST
  PSF map and SDSS shear catalogue. 
\item \emph{SDSS PSF}: To test for artefacts associated with the SDSS
  PSF, we construct SDSS PSF maps using the quoted values in the SDSS
  survey for the reconstructed PSF at the location of each
  galaxy. The resulting PSF ellipticity maps are then cross-correlated
  with the FIRST shear catalogue. 
\item \emph{FIRST P(S) $>$ 0.05}: This final test is not strictly a
  null test. However, we implement it to test for any dependence of
  our measurement on the likelihood of the FIRST sources being
  sidelobe residuals. To perform this test, we create an alternate
  FIRST shape catalogue based on only those sources that have a higher
  chance of being a sidelobe, $P(S)>0.05$. (Recall our main analysis
  is based on only sources with $P(S) < 0.05$.) We then repeat the
  FIRST-SDSS cross-power spectrum analysis using the $P(S) > 0.5$
  alternative FIRST catalogue.
\end{enumerate}

\begin{figure*}
\subfigure{\includegraphics[width=5.76cm]{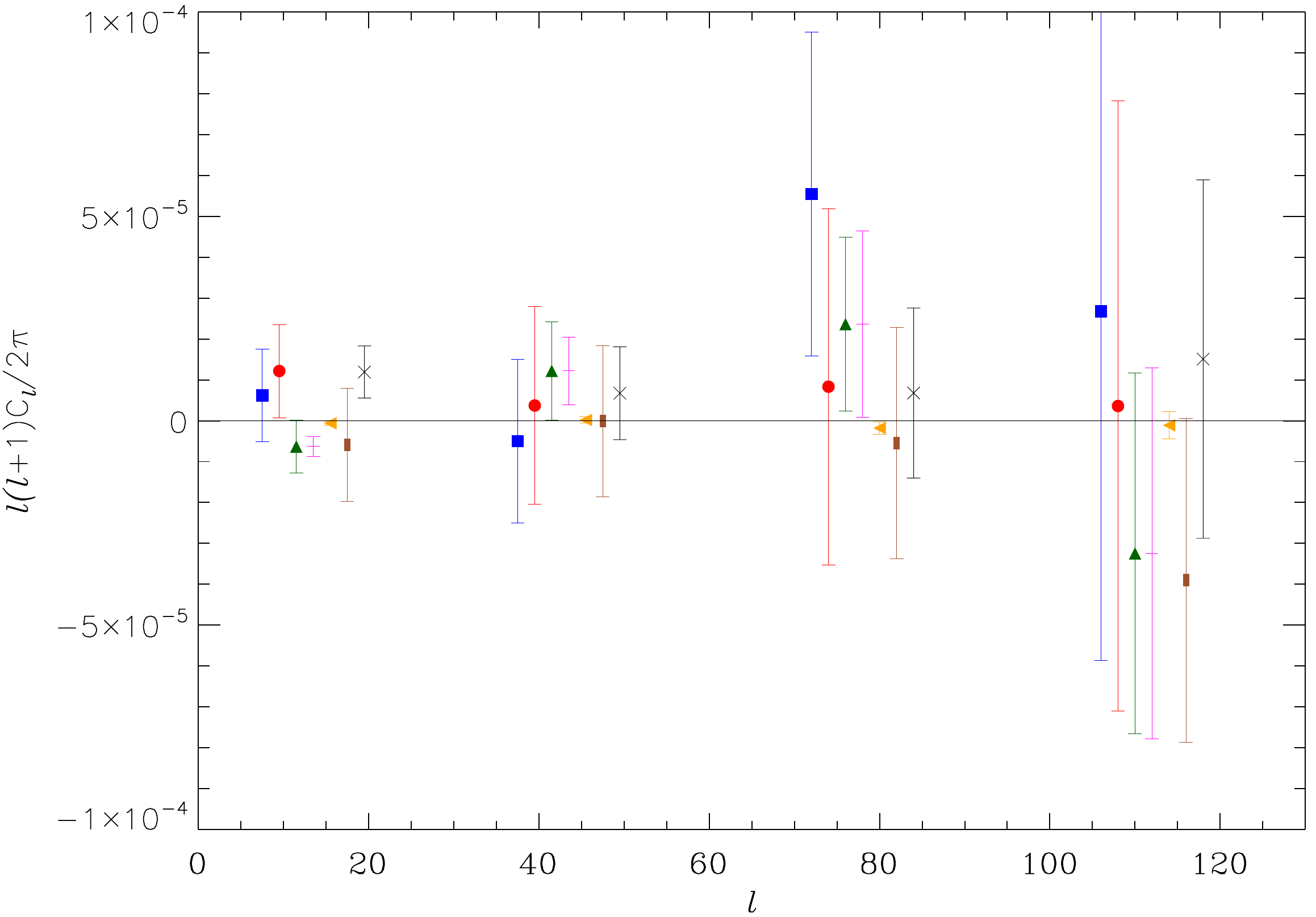}}\hspace{0.5em}
\subfigure{\includegraphics[width=5.76cm]{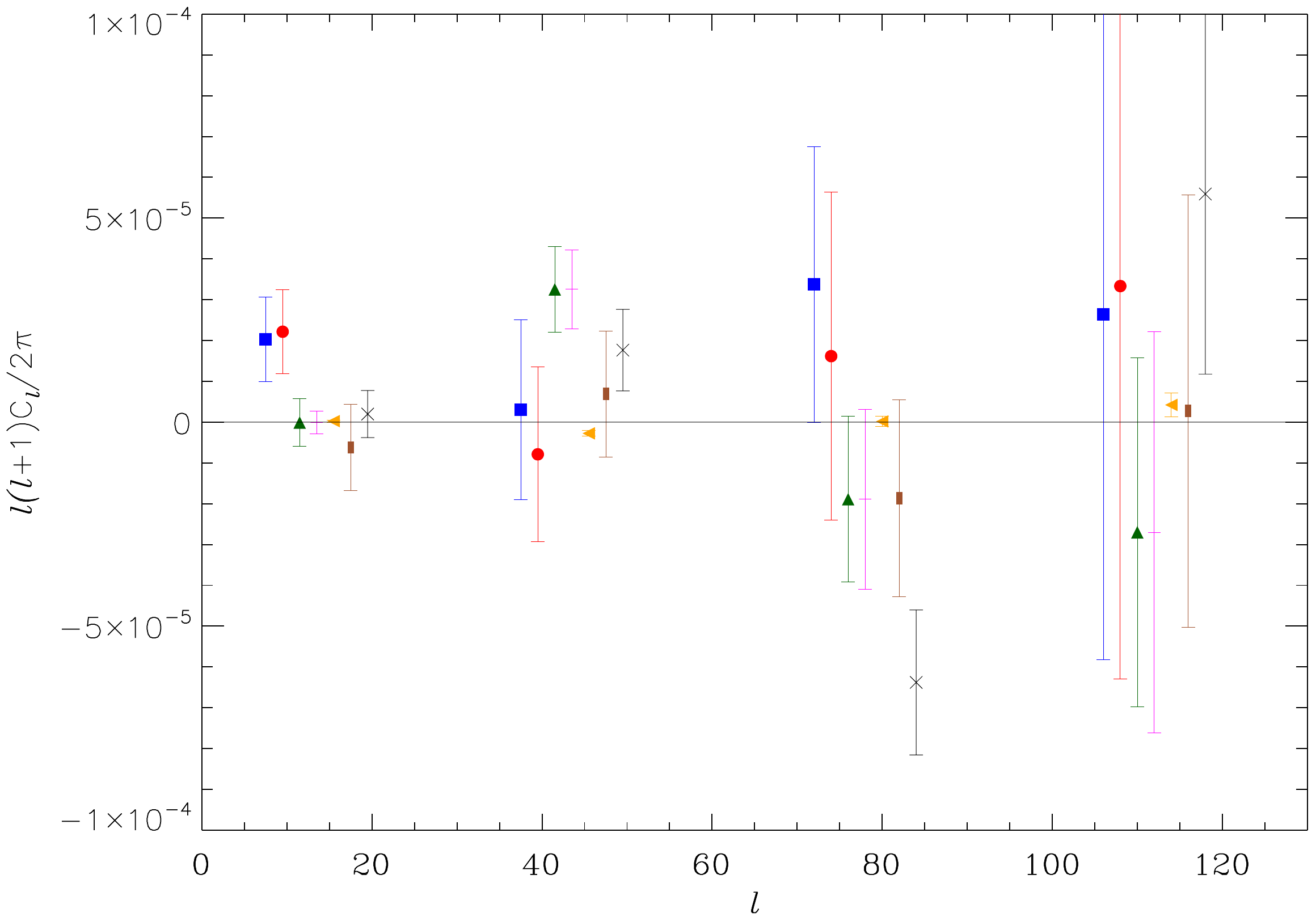}}\hspace{0.5em}
\subfigure{\includegraphics[width=5.76cm]{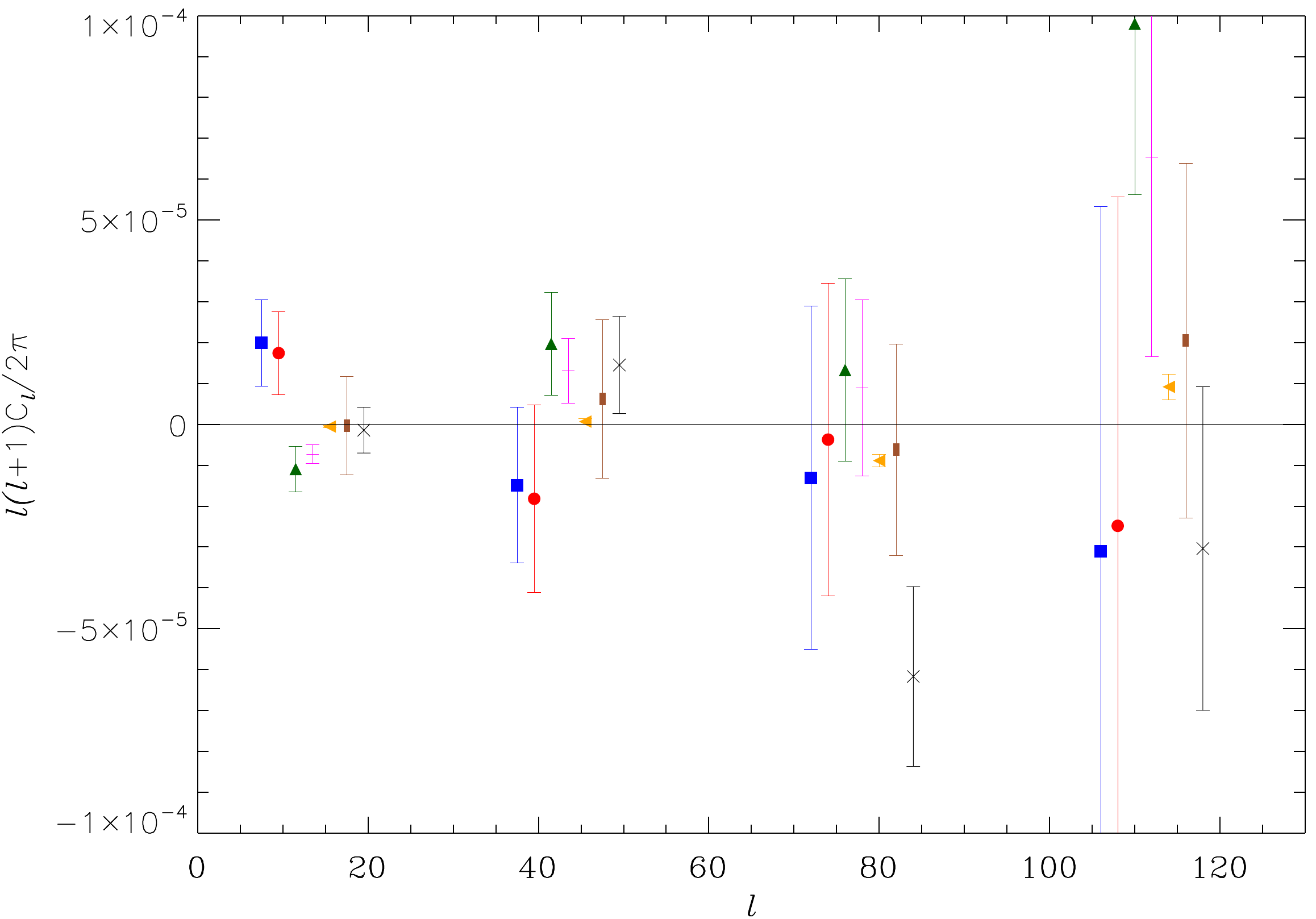}}\hspace{0.5em}
\caption{The $C_\ell^{\kappa\kappa}$ (\emph{left}), $C_\ell^{\beta\beta}$ (\emph{centre})
  and $C_\ell^{\kappa\beta}$ (\emph{right}) power spectra measurements
  for the null tests described in Section~\ref{sec:null_tests}. The
  displayed results are as follows: \emph{North-South}: blue squares;
  \emph{East-West}: red circles; \emph{FIRST random}: green face up
  triangles; \emph{SDSS random}: magenta horizontal lines; \emph{FIRST
    PSF}: orange face left triangles; \emph{SDSS PSF}: brown
  rectangles; \emph{FIRST P(S)$>$0.05}: black crosses.}
\label{fig:nulltests}
\end{figure*}

The $C_\ell^{\kappa\kappa}$, $C_\ell^{\beta\beta}$ and
$C_\ell^{\kappa\beta}$ power spectra measured for all of these tests
are shown in Fig.~\ref{fig:nulltests}. In order to interpret these
null tests, in Fig.~\ref{fig:null_chi2}, we once again plot the
$\chi^2$ measurement from the real data alongside the $\chi^2$
histograms from the simulations. When calculating the $\chi^2$ values
for the null tests, we set $P^{\rm th}_b = 0$ in
equation~(\ref{eq:chi2}). Examination of Figs.~\ref{fig:nulltests} and
\ref{fig:null_chi2} show that the data passes most of these null tests
in the sense that the $\chi^2$ values from the real data are usually
consistent with being a random sample of the simulation $\chi^2$
distribution. Potential problem cases are the \emph{East-West}
$C_\ell^{\beta\beta}$ and \emph{SDSS PSF} $C_\ell^{\kappa\beta}$ null
tests. However, taken as a population, these tests do not reveal any
major outstanding problems in the cross-power spectrum results. To
further quantify the results of these tests, in
Table~\ref{tab:pte_table}, we list the probability to exceed (PTE)
values for each test, which gives the probability that the $\chi^2$
value measured from the real data is consistent with being a random
sampling of the simulation distribution. Low values of these PTE
numbers indicate potential residual systematics. For completeness, in
Fig.~\ref{fig:null_chi}, we also present the measurements of $\chi$
(equation~\ref{eq:chi} with $P_b^{\rm th} = 0$) from the real data
along with the simulation histograms. These tests reveal no obvious
problems with the analysis.
\begin{table}
\caption{PTE values from the the $\chi^2$ null tests described in Section~\ref{sec:null_tests}.}
\begin{center}
\begin{tabular}{c|r|r|r}
Null test & $C_\ell^{\kappa\kappa}$ & $C_\ell^{\beta\beta}$ & $C_\ell^{\kappa\beta}$ \\
\hline
\emph{North-South}:       & 0.64 & 0.10 & 0.10 \\
\emph{East-West}:         & 0.81 & 0.01 & 0.12 \\
\emph{FIRST random}:      & 0.41 & 0.53 & 0.52 \\
\emph{SDSS random}:       & 0.35 & 0.10 & 0.19 \\
\emph{FIRST PSF}:         & 0.96 & 0.85 & 0.87 \\
\emph{SDSS PSF}:          & 0.18 & 0.67 & 0.02 \\
\emph{FIRST P(S)$>$0.05}: & 0.35 & 0.31 & 0.99 \\
\hline
\end{tabular}
\end{center}
\label{tab:pte_table}
\end{table}

\subsection{Constraints on the FIRST and SDSS median redshifts,
  $\sigma_8$ and $\Omega_m$}
Having demonstrated the validity of our results using null tests, we
now use our cross-power spectrum measurement to constrain the median
redshifts of the two populations, and the power spectrum
normalisation, $\sigma_8$ and matter density, $\Omega_m$.

To constrain the median redshifts of FIRST and SDSS we perform a
grid-based likelihood analysis and generate the
$C_\ell^{\kappa\kappa}$ spectrum for two populations with median
redshifts in the range $0.05<z^{\rm SDSS}_m<5.05$ and
$0.5<z^{\rm FIRST}_m<5.5$. The grid resolution in both directions is
$\Delta_z=0.01$. We fix the cosmological parameters values at $\sigma_8=0.8347$
and $\Omega_m=0.3175$ \citep{planck2014}. The model FIRST-SDSS
cross-power spectrum is generated using equation~(\ref{eq:clkk_theory}) and is
then averaged in band powers similarly to what was done during the
power spectrum estimation. 

\begin{figure} 
\centering \includegraphics[width=8.5cm]{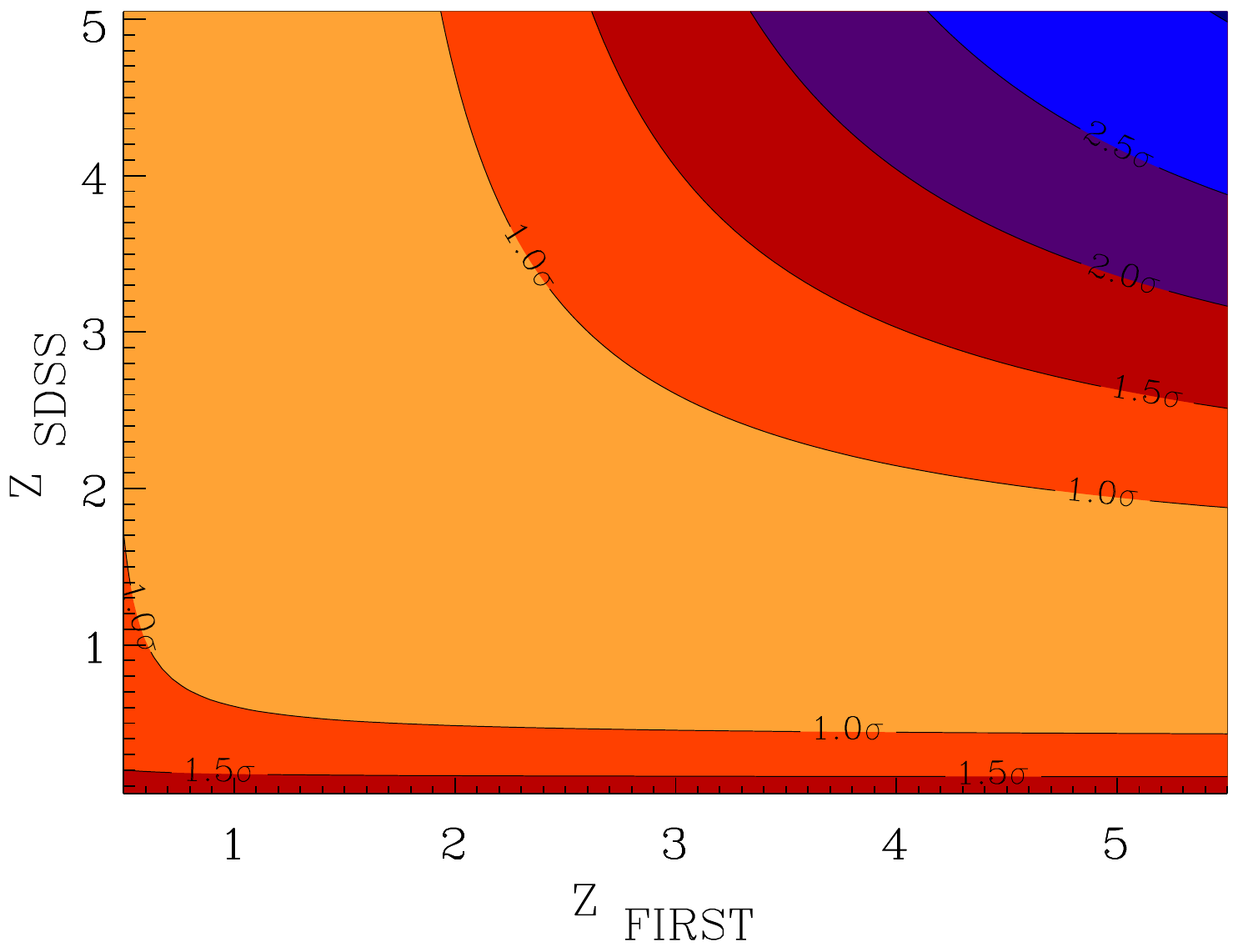}
\caption{Joint constraints on the median redshifts of the SDSS and
  FIRST surveys obtained from fitting theoretical models to our
  $C_\ell^{\kappa\kappa}$ cross-power spectrum
  measurements. Cosmological parameters were kept fixed at the
  concordance values reported in \citet{planck2014}.}
\label{fig:zzplot}
\end{figure}

For each point in parameter space, we construct the $\chi^2$ misfit
statistic (equation~\ref{eq:chi2}) which we then convert to likelihood
values for a model with two degrees of freedom. The results are shown
in Fig.~\ref{fig:zzplot}. We immediately see that the constrains are
symmetric about the $z^{\rm SDSS}_m = z^{\rm FIRST}_m$ line. However,
for the purposes of this study we will make the reasonable assumption
that $z^{\rm SDSS}_m < z^{\rm FIRST}_m$. With this prior in place, the
best fitting values for the median redshifts are $z^{\rm SDSS}_m=1.5$
and $z^{\rm FIRST}_m=1.75$. However, the data are also entirely
consistent (within 1\,$\sigma$) with the median redshifts derived from
the literature of $z^{\rm SDSS}_m=0.53$ \citep{sypniewski2014} and
$z^{\rm FIRST}_m=1.2$ \citep{wilman2008}.

\begin{figure} 
\centering
\includegraphics[width=8.5cm]{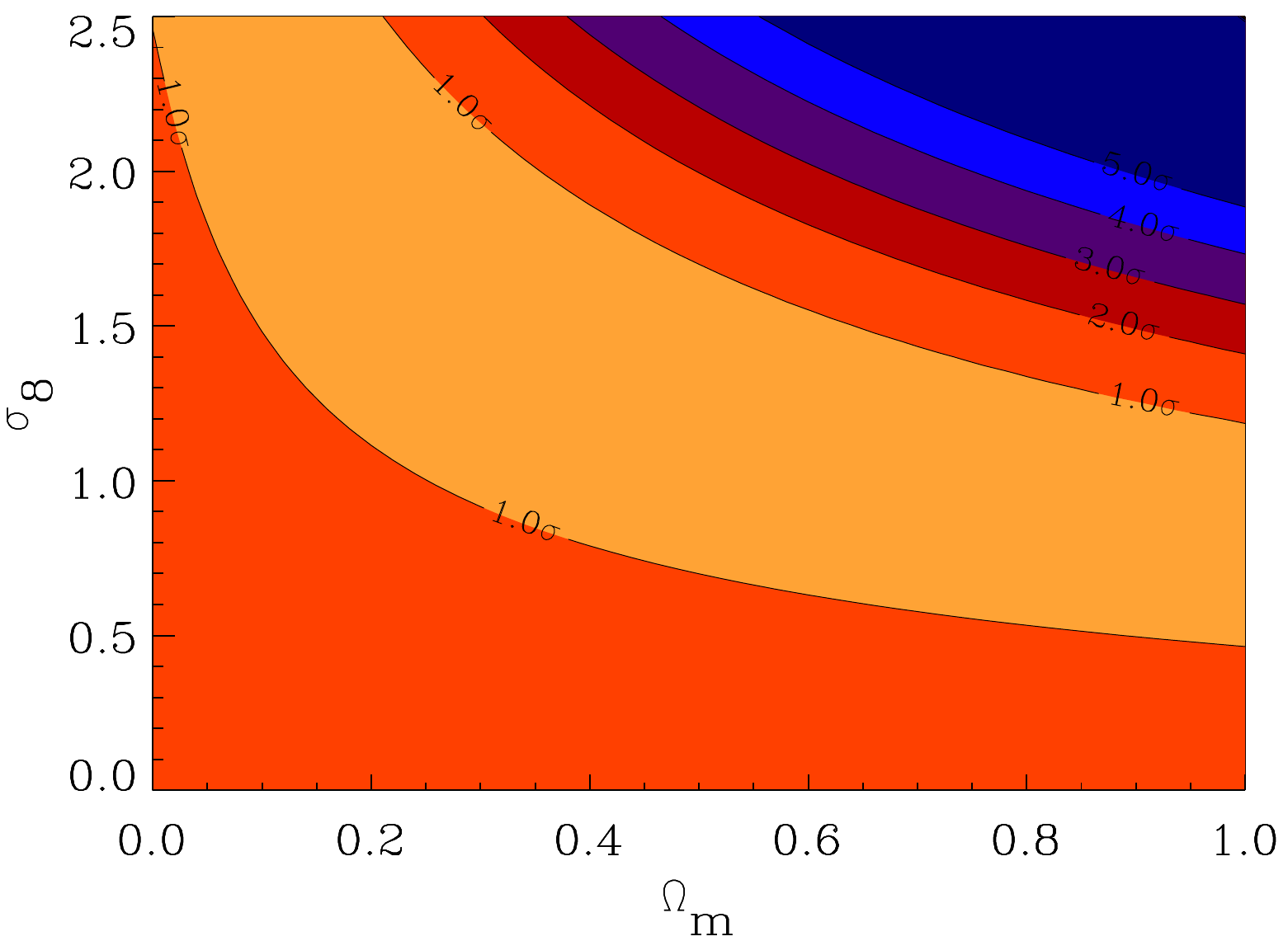} 
\caption{Joint constraints on the matter density, $\Omega_m$ and power
  spectrum normalisation, $\sigma_8$ from fitting theoretical models to our
  $C_\ell^{\kappa\kappa}$ cross-power spectrum measurements. The
  median redshifts were fixed at $z_m^{\rm SDSS}=0.53$ and
  $z_m^{\rm FIRST}=1.2$.}
\label{fig:sigma8omegamatter}
\end{figure}

We now fix the median redshifts for SDSS and FIRST to the literature
values ($z^{\rm SDSS}_m$=0.53 and $z^{\rm FIRST}_m$=1.2) and use our
$C_\ell^{\kappa\kappa}$ measurements to place constraints on the
cosmological parameters, $\sigma_8$ and $\Omega_m$. The results are
shown in Fig.~\ref{fig:sigma8omegamatter}. The best fitting
values for the two parameters are $\sigma_8=1.5^{+0.6}_{-0.8}$
(68$\%$) and $\Omega_m=0.3^{+0.3}_{-0.2}$ (68$\%$). Our results are in
agreement with the values in \citet{planck2014} at the 1$\sigma$
level. Our results are also in good agreement with the $\sigma_8$
constraint obtained from the FIRST cosmic shear analysis of
\citet{chang2004}.

\section{Conclusions}\label{sec:concl}
We have presented a cosmic shear cross-power spectrum analysis of the
SDSS and FIRST surveys, a pair of optical and radio sky surveys with
approximately 10,000 deg$^2$ of overlapping sky coverage. 

The motivation for our study has been to demonstrate the power of
optical-radio cross-correlation analyses for mitigating systematic
effects in cosmic shear analyses. The shear maps that we have
constructed from both the SDSS and the FIRST catalogues are severely
affected by systematic effects. Measuring the auto-shear power
spectrum from either of the datasets results in a heavily biased
measurement -- in both cases, the measured convergence power spectrum,
$C_\ell^{\kappa\kappa}$ is more than an order of magnitude larger than
that expected in the concordance cosmological model. The presence of a
large $B$-mode signal ($C_\ell^{\beta\beta}$) in both the SDSS and
FIRST auto power spectra is further evidence that the shear maps are
contaminated by large-scale systematic effects. For our SDSS maps,
these large-scale systematics dominate over the random noise on the
scales of interest here ( $\ell < 120$). For FIRST, the spurious power
due to the large-scale systematics is approximately the same as the
noise power spectrum. Note however that the random noise component in
the FIRST shear maps is much larger than in the SDSS maps due to the
much small galaxy number density in FIRST.
Despite these very large shear systematics, we have demonstrated using
simulations that one can still recover a cosmic shear signal using the
FIRST-SDSS cross-power spectrum. Although the cross-power spectrum is
not affected by uncorrelated systematics in the mean, the associated
uncertainties on the cross-power spectrum measurements are amplified
by the presence of the FIRST and SDSS systematic effects. In our final
analysis, our total errors on $C_\ell^{\kappa\kappa}$ have increased by
a factor of $\sim\!\!2.5$ compared to those due to cosmic variance and
random noise alone. Although not the focus of this paper, it is likely
that this enhancement in the cross-power spectrum errors could be
significantly reduced by applying more sophisticated techniques for
correcting the galaxy shapes measurements for the effects of PSF
anisotropy in both surveys, prior to construction of the final shear
maps.

By cross-correlating the SDSS and the FIRST data, we tentatively
detect a signal in the $C_\ell^{\kappa\kappa}$ power spectrum that is
inconsistent with zero at the 99\% confidence level. In contrast to
almost all cosmic shear analyses to date (which primarily probe the
shear signal on sub-degree scales), our measurements constrain the
weak lensing power spectrum on large angular scales. Our measurements
probe the power spectrum in the fully linear r\'egime, in the
multipole range $10 \lsim \ell \lsim 100$, corresponding to angular
scales $2^\circ \lsim \theta \lsim 20^\circ$. 

Our results are consistent (within $\sim\!\!1\sigma$) with the expected
signal in the concordance cosmological model, assuming median
redshifts of $z_m^{\rm SDSS}=0.53$ and $z_m^{\rm FIRST}=1.2$ for the
SDSS and FIRST surveys respectively. Our measurements of the
odd-parity ($C_\ell^{\beta\beta}$) and parity-violating
($C_\ell^{\kappa\beta}$) cross-power spectra are fully consistent with
zero, which demonstrates the success of the cross-power spectrum
approach in mitigating systematic effects. We have also validated our
analysis and results by performing a range of null tests on the data.

We have used our $C_\ell^{\kappa\kappa}$ measurement to jointly
constrain the median redshifts of the two surveys. With cosmological
parameters fixed at their concordance values and assuming $z_m^{\rm
  SDSS} < z_m^{\rm FIRST}$, we find best fitting value of $z_m^{\rm
  SDSS} = 1.5$ and $z_m^{\rm FIRST} = 1.75$. However, our constraints
are weak and our measurements are also consistent with the literature
values of $z_m^{\rm SDSS} = 0.53$ and $z_m^{\rm FIRST} = 1.2$. Fixing
the median redshifts to these literature values, we have used our
measurements to constrain the cosmological parameters $\Omega_m$ and
$\sigma_8$, where we find best fitting values of
$\Omega_m=0.30^{+0.3}_{-0.2}$ and $\sigma_8=1.5^{+0.6}_{-0.8}$ (68\%
confidence levels), which are consistent with values quoted in
\citet{planck2014} at the 1$\sigma$ level. 

Although the detection of $C_\ell^{\kappa\kappa}$ presented here is
tentative and lacks the precision acheived by state-of-the-art optical
weak lensing surveys, we believe the analysis techniques developed in
this paper will prove extremely useful for future high precision
cosmic shear analyses. In particular, we have successfully
demonstrated that one can extract an unbiased cosmic shear signal even
in the presence of severe shear systematics using the cross-power
spectrum approach. This type of analysis will be well suited for
performing cross-correlation studies of future overlapping optical and
radio surveys (e.g. with SKA, LSST and \emph{Euclid}). Cosmic shear
analyses of these future surveys will require exquisite control of
systematic effects if they are to meet their science goals of per
cent level constraints on dark energy and modified gravity
thoeries. The cross-power spectrum approach that we have demonstrated
for the first time in this paper represents a very promising tool for
achieveing the required control of systematic effects.

\section*{Acknowledgments}
We thank Neal Jackson, Ian Harrison, Nick Wrigley and Stuart Harper for useful
discussions. CD acknowledges the support of a STFC quota studentship
and a President's Doctoral Scholarship from the University of
Manchester. CD and MLB are grateful to the European Research Council
for support through the award of an ERC Starting Independent
Researcher Grant (EC FP7 grant number 280127). MLB also thanks the
STFC for the award of Advanced and Halliday fellowships (grant number
ST/I005129/1).

Funding for SDSS-III has been provided by the Alfred P. Sloan
Foundation, the Participating Institutions, the National Science
Foundation, and the U.S. Department of Energy Office of Science. The
SDSS-III web site is http://www.sdss3.org/.

SDSS-III is managed by the Astrophysical Research Consortium for the
Participating Institutions of the SDSS-III Collaboration including the
University of Arizona, the Brazilian Participation Group, Brookhaven
National Laboratory, Carnegie Mellon University, University of
Florida, the French Participation Group, the German Participation
Group, Harvard University, the Instituto de Astrofisica de Canarias,
the Michigan State/Notre Dame/JINA Participation Group, Johns Hopkins
University, Lawrence Berkeley National Laboratory, Max Planck
Institute for Astrophysics, Max Planck Institute for Extraterrestrial
Physics, New Mexico State University, New York University, Ohio State
University, Pennsylvania State University, University of Portsmouth,
Princeton University, the Spanish Participation Group, University of
Tokyo, University of Utah, Vanderbilt University, University of
Virginia, University of Washington, and Yale University.

Some of the results in this paper have been derived using
the {\sevensize HEALPix}~\citep{gorski2005} package.

\bibliographystyle{mn2e_plus_arxiv}
\bibliography{powerspectrum}

\appendix
\section{Results from null tests}
\begin{figure*}
\centering
\includegraphics[width=7in,height=9in]{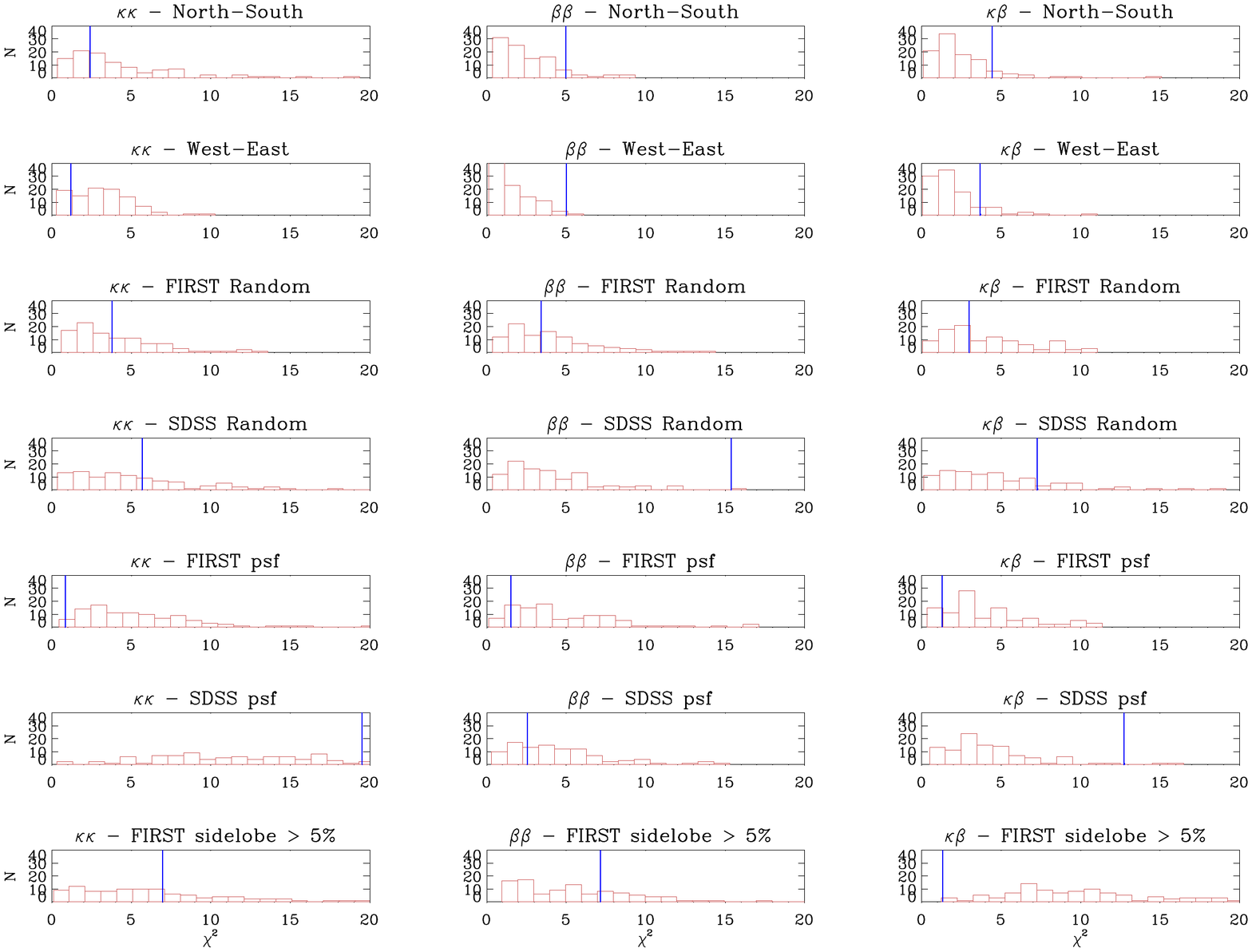} 
\caption{Histograms showing the distribution of $\chi^2$ values (equation~\ref{eq:chi2})
  measured from the simulations for the suite of null tests described
  in Section~\ref{sec:null_tests}. The results are shown for the
  $C_\ell^{\kappa\kappa}$ (\emph{left column}), $C_\ell^{\beta\beta}$
  (\emph{centre column}) and $C_\ell^{\kappa\beta}$ (\emph{right
    column}) power spectra. Over-plotted as the vertiocal blue line is
  the equivalent value for the real data measurement.}
\label{fig:null_chi2}
\end{figure*}

\begin{figure*}
\centering
\includegraphics[width=7in,height=9in]{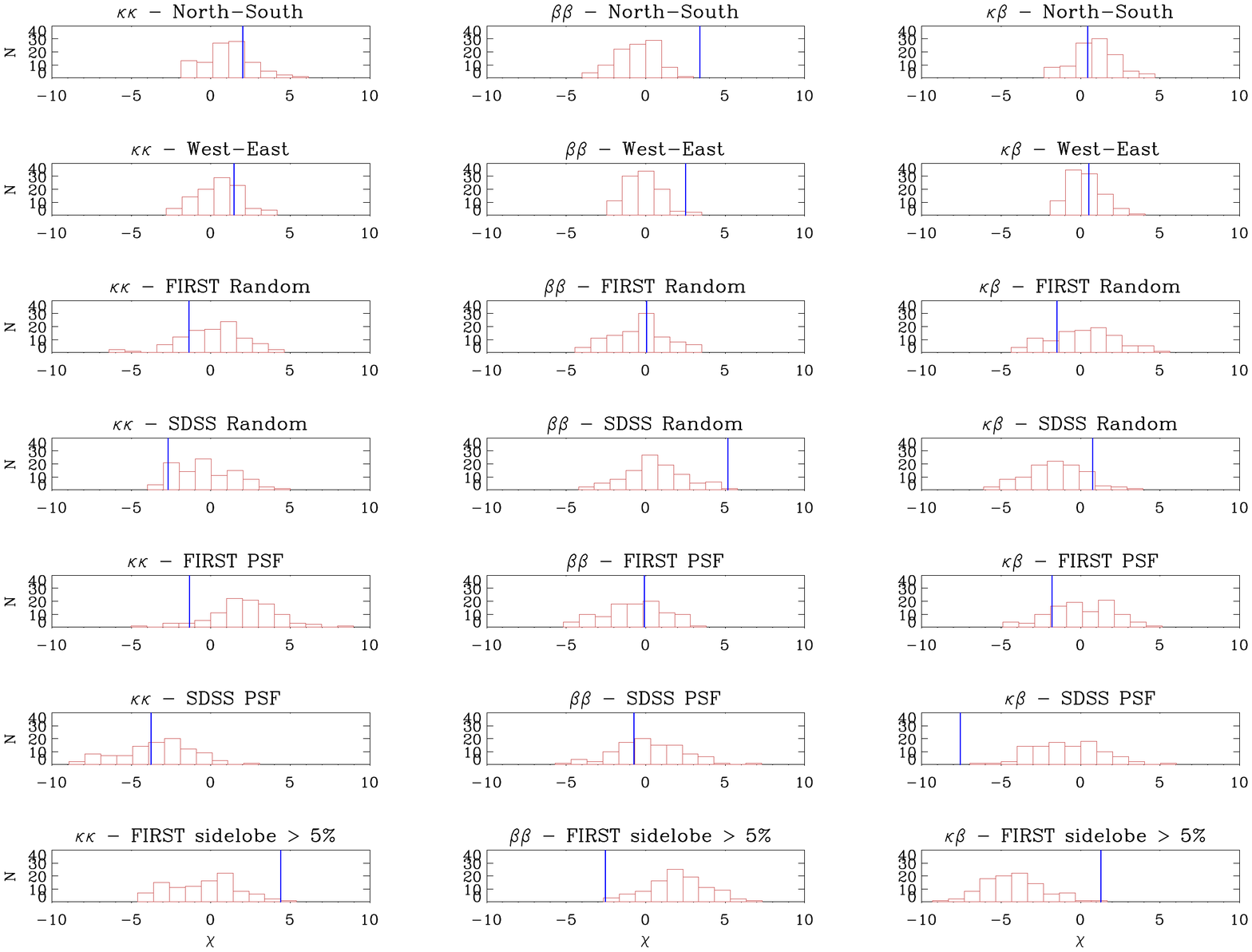} 
\caption{Histograms showing the distribution of $\chi$ values (equation~\ref{eq:chi})
  measured from the simulations for the suite of null tests described
  in Section~\ref{sec:null_tests}. The results are shown for the
  $C_\ell^{\kappa\kappa}$ (\emph{left column}), $C_\ell^{\beta\beta}$
  (\emph{centre column}) and $C_\ell^{\kappa\beta}$ (\emph{right
    column}) power spectra. Over-plotted as the vertiocal blue line is
  the equivalent value for the real data measurement.}
\label{fig:null_chi}
\label{lastpage}
\end{figure*}

Fig.~\ref{fig:null_chi2} compares the $\chi^2$
statistic (equation~\ref{eq:chi2})) for the data with the distribution of
$\chi^2$ measured from the simulations for the suite the null tests
described in Section~\ref{sec:null_tests}. Fig.~\ref{fig:null_chi}
shows the same comparison for the $\chi$ statistic of
equation~(\ref{eq:chi}). 

\end{document}